\documentclass[sigconf]{acmart}
\usepackage{balance}

\usepackage{microtype}

\usepackage{booktabs}
\usepackage{xspace}
\usepackage{hyperref}
\usepackage{graphicx}
\usepackage{multirow}
\usepackage{listings}
\usepackage{makecell}
\usepackage{enumitem}
\usepackage{tikz}

\definecolor{gray}{RGB}{211,211,211}
\definecolor{javared}{rgb}{0.6,0,0} 
\definecolor{javagreen}{rgb}{0.25,0.5,0.35} 
\definecolor{javapurple}{rgb}{0.5,0,0.2} 
\definecolor{javadocblue}{rgb}{0.25,0.35,0.75} 
\newcommand{\jbasicstyle}{\small\sffamily} 

\newcommand{\jnumberstyle}{\scriptsize}

\lstdefinelanguage{pseudo}
{
  morekeywords={},
  keywordstyle=\bfseries,
  lineskip=-0.1em,
  numbers=left, 
  numberstyle=\jnumberstyle,
  numbersep=4pt,
  basicstyle=\jbasicstyle,
  breaklines=true,
  breakautoindent=true,
  tabsize=2,
  columns=fullflexible,
  morecomment=*[l][\textsl]{//},
  mathescape=true,
  xleftmargin=10pt,
}

\lstdefinelanguage{todo-comment}
{
  morekeywords={},
  keywordstyle=\bfseries,
  lineskip=-0.1em,
  numbers=none,
  basicstyle=\jbasicstyle,
  breaklines=true,
  breakautoindent=true,
  tabsize=2,
  columns=fullflexible,
  morecomment=*[l][\textsl]{//},
  mathescape=true,
  xleftmargin=-10pt,
}

\lstdefinelanguage{java-pretty}
{
  language=java,
  numbers=left,
  basicstyle=\scriptsize\ttfamily,
  numberstyle=\scriptsize,
  breaklines=true,
  columns=fullflexible,
  xleftmargin=16pt,
  showstringspaces=false,
  keywordstyle=\color{javapurple}\bfseries,
  stringstyle=\color{javared},
  commentstyle=\color{javagreen},
  morecomment=[s][\color{javadocblue}]{/**}{*/},
}

\usetikzlibrary{shapes,arrows,positioning,calc}

\tikzstyle{text_box} = [rectangle, text width=15em, text centered]
\tikzstyle{block} = [draw=black, thick, rectangle, text=black, text width=8em, text centered, rounded corners, minimum height=7ex] 
\tikzstyle{line} = [draw, line width=0.25mm, -latex']
\tikzstyle{long_text_box} = [rectangle, text width=15em, minimum size=2cm]
\newcommand*\circled[1]{\tikz[baseline=(char.base)]{
    \node[shape=circle, draw=black, minimum size=0.1, inner sep=0.8pt, fill=white, thick, text=black] (char) {#1};}}
\newcommand{\XSpace}[1]{}
\newcommand{\XComment}[1]{}

\newcommand{\DefMacro}[2]{\expandafter\newcommand\csname rmk-#1\endcsname{#2}}
\newcommand{\UseMacro}[1]{\csname rmk-#1\endcsname}
\newcommand{\RQ}[2]{\noindent\textbf{RQ#1:} #2}

\newcommand{\MyPara}[1]{\vspace{2pt}\noindent\textbf{#1}.}

\newcommand{\InputWithSpace}[1]{\bgroup\def\arraystretch{1.15}\input{#1}\egroup}
\newcommand{\Code}[1]{{\ifmmode{\mathtt{#1}}\else$\mathtt{#1}$\fi}}
\newcommand{\CodeIn}[1]{{\ifmmode{\mathtt{#1}}\else$\mathtt{#1}$\fi}}

\usepackage{mdwlist}

\newcommand{\MOEDIT}{\textsc{CoditT5}\xspace}
\newcommand{\MOEDITft}{CodeT5 (w/ edit-based output)\xspace}
\newcommand{\Title}{\MOEDIT: \Pretraining for Source Code and \\ Natural Language Editing}
\newcommand{\ShortTitle}{\MOEDIT: \Pretraining for Source Code and Natural Language Editing}

\newcommand{\HighestCopyRate}{34.25\%\xspace}

\DefMacro{FCap-example-moedit-better}{An example in \codereview task where \PLBARTft merely copies
  the input which does not match reviewer's comment.}
\DefMacro{FCap-example-plbart-better}{An example of \PLBART generating
  better output than \MOEDIT, in \codereview task.  \MOEDIT results in
  wrong number of arguments for method \CodeIn{select}.}
\DefMacro{FCap-pretrain-model}{The corrupted text is
  encoded with\XComment{ \PLBART} a bidirectional encoder, and the
  decoder is \pretrained to generate \editseqs to recover the original
  text followed by a separation token (<s>), and finally the target sequence}
\DefMacro{FCap-example-new-objective}{Example of new objective}

\newcommand{\editseqs}{sequences of edit actions\xspace}

\newcommand{\pretrain}{pretrain\xspace}

\newcommand{\pretrained}{pretrained\xspace}
\newcommand{\Pretrained}{Pretrained\xspace}
\newcommand{\pretraining}{pretraining\xspace}
\newcommand{\Pretraining}{Pretraining\xspace}
\newcommand{\finetune}{fine-tune\xspace}

\newcommand{\finetuned}{fine-tuned\xspace}

\newcommand{\finetuning}{fine-tuning\xspace}
\newcommand{\Finetuning}{Fine-tuning\xspace}

\DefMacro{PLBART}{\textbf{PLBART-finetuned}}
\DefMacro{PLEDIT}{\textbf{PLEDIT-wo-pretrain}}
\DefMacro{MOEDIT}{\textbf{PLEDIT}}
\DefMacro{MOEDIT-rerank}{\textbf{PLEDIT-Rerank}}

\newcommand{\editmodel}{\citet{PanthaplackelETAL20Learning}\xspace}
\newcommand{\CodeTf}{CodeT5\xspace}
\newcommand{\PLBART}{\CodeTf}
\newcommand{\PLBARTft}{PLBART\xspace}

\newcommand{\MODIT}{MODIT\xspace}
\newcommand{\PLBARTVa}{PLBART\xspace}
\newcommand{\MOEDITVa}{PLBART-fe\xspace}
\newcommand{\MOEDITRerank}{\MOEDIT (reranked with CodeT5)\xspace}
\newcommand{\PLBARTRerank}{CodeT5 (reranked with \MOEDIT)\xspace}
\newcommand{\PLBARTCodeTRerank}{PLBART-CodeT5-rerank\xspace}
\newcommand{\EditModel}{\citet{PanthaplackelETAL20Learning}\xspace}

\newcommand{\inp}{\textit{inp}\xspace}
\newcommand{\out}{\textit{out}\xspace}
\newcommand{\noisyinp}{\textit{inp$'$}\xspace}

\newcommand{\bfs}{$B2F_{s}$\xspace}
\newcommand{\bfm}{$B2F_{m}$\xspace}

\newcommand{\xMatch}{\textbf{xMatch}\xspace}

\newcommand{\bugfix}{bug fixing\xspace}
\newcommand{\codereview}{automated code review\xspace}
\newcommand{\cupdate}{comment updating\xspace}

\newcommand{\Bugfix}{Bug Fixing\xspace}
\newcommand{\Codereview}{Automated Code Review\xspace}
\newcommand{\Cupdate}{Comment Updating\xspace}

\newcommand{\Insert}{\texttt{Insert}\xspace}
\newcommand{\Delete}{\texttt{Delete}\xspace}

\newcommand{\introReviewerComment}{Generally better to qualify than making static import\xspace}

\DefMacro{THead-train}{\Train\xspace}
\DefMacro{THead-valid}{\Val\xspace}
\DefMacro{THead-test}{\Test\xspace}
\DefMacro{THead-all}{All\xspace}

\DefMacro{THead-task}{Task\xspace}
\DefMacro{THead-dataset}{Dataset\xspace}
\DefMacro{THead-train-examples-count}{Train\xspace}
\DefMacro{THead-valid-examples-count}{Valid\xspace}
\DefMacro{THead-test-examples-count}{Test\xspace}
\DefMacro{ds-comment-update-dataset}{Clean Dataset\xspace}
\DefMacro{ds-comment-update-full-dataset}{Full Dataset\xspace}
\DefMacro{ds-bf-small-dataset}{\bfs\xspace}
\DefMacro{ds-bf-medium-dataset}{\bfm\xspace}
\DefMacro{ds-code-review-dataset}{Code Review\xspace}

\DefMacro{TCap-downstream-dataset}{Statistics of downstream dataset.}
\DefMacro{TCap-models-copy-pct-results}{Percentage that model just copy the input.}
\DefMacro{TCap-plan-edit-consistent}{Percentage that edit plan is consistent with target output.}
\newcommand{\Train}{Training\xspace}
\newcommand{\Val}{Validation\xspace}
\newcommand{\Test}{Test\xspace}

\DefMacro{THead-ours}{\MOEDIT}
\DefMacro{THead-ours-rerank}{\MOEDITRerank}
\DefMacro{THead-MOEDIT-rerank}{\MOEDITRerank}
\DefMacro{THead-modit}{\MODIT}
\DefMacro{THead-EditModel}{\EditModel}
\DefMacro{THead-models}{Models}
\DefMacro{THead-plbart}{\PLBARTft}
\DefMacro{THead-T5}{T5}
\DefMacro{THead-plbart-no-finetune}{\PLBARTVa}
\DefMacro{THead-PLEDIT}{\MOEDITVa}
\DefMacro{THead-CodeT5}{CodeT5}
\DefMacro{THead-CoditT5-ft}{CodeT5 (w/ edit-based output)}
\DefMacro{THead-CoditT5}{CoditT5}
\DefMacro{THead-PLBART-rerank}{\PLBARTRerank}
\DefMacro{THead-PLBART-CodeT5-rerank}{\PLBARTCodeTRerank}
\DefMacro{THead-CodeT5-EditT5-rerank}{CodeT5-EditT5-rerank}
\DefMacro{THead-EditT5-CodeT5-rerank}{EditT5-CodeT5-rerank}
\DefMacro{THead-EditT5}{EditT5}
\DefMacro{THead-CodeT5-plbart-rerank}{CodeT5-PLBART-rerank}
\DefMacro{THead-CodeT5-MOEDIT-rerank}{CodeT5-PLEDIT-rerank}
\DefMacro{THead-EditT5-plbart-rerank}{EditT5-PLBART-rerank}
\DefMacro{THead-EditT5-MOEDIT-rerank}{EditT5-PLEDIT-rerank}
\DefMacro{THead-CoditT5-CodeT5-rerank}{CoditT5 (reranked with CodeT5)}
\DefMacro{THead-CodeT5-CoditT5-rerank}{CodeT5 (reranked with CoditT5)}
\DefMacro{THead-EditT5-dp}{EditT5 (domain pretrain)}

\DefMacro{THead-Acc-top-1}{xMatch}
\DefMacro{THead-BLEU}{BLEU-4}
\DefMacro{THead-SARI}{SARI}
\DefMacro{THead-GLEU}{GLEU}
\DefMacro{THead-CodeBLEU}{CodeBLEU}
\DefMacro{THead-METEOR}{METEOR}
\DefMacro{THead-examples-count}{Examples}
\DefMacro{THead-avg-source-token-count}{Avg. source tokens}
\DefMacro{THead-avg-context-token-count}{Avg. context tokens}
\DefMacro{THead-avg-target-token-count}{Avg. target tokens}
\DefMacro{THead-avg-edit-token-count}{Avg. target edit tokens}
\DefMacro{THead-avg-lev-distance}{Avg. lev. distance}

\DefMacro{TCap-pretrain-dataset}{Statistics of the datasets used to
  \pretrain \MOEDIT. First row: number of programming language and natural language; second row:
  average number of tokens in corrupted input sequences\XComment{ after sentencepiece
    tokenization}; third row: average number of tokens in the output sequence (edit plan + target sequence)\XComment{ after
    sentencepiece tokenization}.}
\DefMacro{THead-buggy-avg-tokens-count}{Avg. C-Tokens}
\DefMacro{THead-edit-avg-tokens-count}{Avg. O-Tokens}
\DefMacro{THead-code}{PL}
\DefMacro{THead-nl}{NL}

\DefMacro{TCap-bf-small-dataset}{Statistics of the \bugfix \bfx dataset}
\DefMacro{TCap-bf-medium-dataset}{Statistics of the \bugfix \bfm dataset}

\DefMacro{TCap-bf-small-models-results}{Results for \bugfix on the \bfs dataset. The results with the same prefixes (e.g., $\beta$) are NOT
  statistically significantly different.}
\DefMacro{TCap-bf-medium-models-results}{Results for \bugfix on the \bfm dataset. The results with the same prefixes (e.g., $\beta$) are NOT
  statistically significantly different.}
\DefMacro{TCap-bf-models-results}{Results on \bugfix dataset. The results with the same prefixes (e.g., $\beta$) are NOT
  statistically significantly different.}

\DefMacro{TCap-comment-update-dataset}{Statistics of the \cupdate dataset}
\DefMacro{TCap-comment-update-full-dataset}{Statistics of the \cupdate dataset with full test set}
\DefMacro{TCap-comment-update-models-results}{Results for \cupdate on the clean test set. The results with the same prefixes (e.g., $\beta$) are NOT
  statistically significantly different.}
\DefMacro{TCap-comment-update-full-models-results}{Results for \cupdate on the full test set. The results with the same prefixes (e.g., $\beta$) are NOT
  statistically significantly different.}

\DefMacro{TCap-code-review-dataset}{Statistics of the \codereview dataset}
\DefMacro{TCap-code-review-models-results}{Results for \codereview. The results with the same prefixes (e.g., $\beta$) are NOT
  statistically significantly different.}

\DefMacro{THead-CoditT5-xMatch-pct}{\MOEDIT}
\DefMacro{THead-CodeT5-xMatch-pct}{\CodeTf}
\DefMacro{THead-Both-xMatch-pct}{same}
\DefMacro{TCap-test-models-xMatch-pct-distribution}{\CodeTf and \MOEDIT
  \xMatch confusion matrix on the test sets of the downstream tasks.}
\DefMacro{TLab-test-models-xMatch-pct-distribution}{models-xMatch-distribution}

\DefMacro{TCap-bf-small-valid-xMatch-statistics}{Our model and PLBART xMatch statistics on Bugfix small validation set.}
\DefMacro{TCap-bf-medium-valid-xMatch-statistics}{Our model and PLBART xMatch statistics on Bugfix medium validation set.}
\DefMacro{TCap-comment-update-valid-xMatch-statistics}{Our model and PLBART xMatch statistics on Comment update validation set.}
\DefMacro{TCap-code-review-valid-xMatch-statistics}{Our model and PLBART xMatch statistics on Code review medium validation set.}

\DefMacro{TCap-valid-models-same-output-statistics}{Our model and PLBART same outputs statistics on validation set.}
\DefMacro{THead-PL}{PL}
\DefMacro{THead-NL}{NL}
\DefMacro{THead-delete-prob}{Probability of \texttt{Delete} edit}
\DefMacro{THead-insert-prob}{Probability of \texttt{Insert} edit}
\DefMacro{THead-replace-prob}{Probability of \texttt{Replace} edit}
\DefMacro{THead-mean-span-length}{Avg. No. of Tokens}
\DefMacro{THead-mean-span-count}{Avg. No. of Spans}
\DefMacro{TCap-span-stats}{Statistics collected from downstream tasks for creating \pretraining dataset. \UseMacro{THead-mean-span-length} represents the average number of tokens in each edited span; \UseMacro{THead-mean-span-count} represents the average number of edited spans in each input sequence.}

\DefMacro{THead-PLBART-xMatch}{\# PLBART xMatch}
\DefMacro{THead-PLBART-nMatch}{\# PLBART not Match}
\DefMacro{THead-MOEDIT-xMatch}{\# PLEDIT xMatch}
\DefMacro{THead-MOEDIT-nMatch}{\# PLEDIT not Match}
\DefMacro{THead-MOEDIT-PLBART-same-output-count}{\# Same}
\DefMacro{THead-MOEDIT-PLBART-same-output-xMatch}{\# xMatch}
\DefMacro{THead-MOEDIT-PLBART-same-output-avg-bleu}{BLEU-4}
\DefMacro{THead-MOEDIT-PLBART-same-output-nMatch-avg-bleu}{nMatch BLEU-4}
\DefMacro{THead-MOEDIT-PLBART-diff-output-avg-self-bleu}{Diff self-bleu}
\DefMacro{THead-MOEDIT-PLBART-avg-self-bleu}{self-bleu}
\DefMacro{THead-MOEDIT-PLBART-optimal-avg-bleu}{opt. bleu}
\DefMacro{THead-MOEDIT-avg-bleu}{PLEDIT bleu}
\DefMacro{THead-MOEDIT-PLBART-diff-oneMatch-avg-self-bleu}{Diff Match self-bleu}
\DefMacro{THead-MOEDIT-PLBART-diff-nMatch-avg-self-bleu}{Diff nMatch self-bleu}

\DefMacro{THead-ds-bf-small}{\Bugfix (small)}
\DefMacro{THead-ds-bf-medium}{\Bugfix (medium)}
\DefMacro{THead-ds-comment-update}{\Cupdate}
\DefMacro{THead-ds-code-review}{\Codereview}

\DefMacro{THead-bf-small}{\Bugfix}
\DefMacro{THead-bf-small1}{\bfs}
\DefMacro{THead-bf-medium}{\bfm}
\DefMacro{THead-comment-update}{\Cupdate}
\DefMacro{THead-comment-update-full}{\Cupdate (full)}
\DefMacro{THead-code-review}{\Codereview}

\DefMacro{pl-pretrain-number}{5.9}
\DefMacro{nl-pretrain-number}{1.6}
\DefMacro{pl-pretrain-number-initial}{6.1}
\DefMacro{nl-pretrain-number-initial}{1.9}
\DefMacro{rerank-best-improve}{19.35}

\DefMacro{coditT5-better-than-codeT5}{9}
\DefMacro{codeT5-edit-better-than-codeT5}{7}
\DefMacro{codeT5-better-than-codeT5-edit}{6}
\DefMacro{coditT5-better-than-codeT5-edit}{11}
\DefMacro{total-metrics-number}{14}

\DefMacro{coditT5-consistent-rate-low}{74\%}
\DefMacro{coditT5-consistent-rate-high}{92\%}

\DefMacro{ds-pretrain-code-buggy-avg-tokens-count}{102.01}
\DefMacro{ds-pretrain-code-edit-avg-tokens-count}{120.23}
\DefMacro{ds-pretrain-code-examples-count}{5,956,069}
\DefMacro{ds-pretrain-nl-buggy-avg-tokens-count}{15.42}
\DefMacro{ds-pretrain-nl-edit-avg-tokens-count}{26.57}
\DefMacro{ds-pretrain-nl-examples-count}{1,675,277}

\DefMacro{ds-bf-medium-test-avg-context-token-count}{151.82}
\DefMacro{ds-bf-medium-test-avg-edit-token-count}{95.68}
\DefMacro{ds-bf-medium-test-avg-lev-distance}{128.65}
\DefMacro{ds-bf-medium-test-avg-source-token-count}{79.95}
\DefMacro{ds-bf-medium-test-avg-target-token-count}{75.65}
\DefMacro{ds-bf-medium-test-examples-count}{6,538}
\DefMacro{ds-bf-medium-train-avg-context-token-count}{152.99}
\DefMacro{ds-bf-medium-train-avg-edit-token-count}{95.63}
\DefMacro{ds-bf-medium-train-avg-lev-distance}{127.43}
\DefMacro{ds-bf-medium-train-avg-source-token-count}{80.08}
\DefMacro{ds-bf-medium-train-avg-target-token-count}{75.38}
\DefMacro{ds-bf-medium-train-examples-count}{52,324}
\DefMacro{ds-bf-medium-valid-avg-context-token-count}{153.59}
\DefMacro{ds-bf-medium-valid-avg-edit-token-count}{95.25}
\DefMacro{ds-bf-medium-valid-avg-lev-distance}{127.15}
\DefMacro{ds-bf-medium-valid-avg-source-token-count}{79.72}
\DefMacro{ds-bf-medium-valid-avg-target-token-count}{75.00}
\DefMacro{ds-bf-medium-valid-examples-count}{6,542}

\DefMacro{ds-bf-small-test-avg-context-token-count}{76.39}
\DefMacro{ds-bf-small-test-avg-edit-token-count}{52.38}
\DefMacro{ds-bf-small-test-avg-lev-distance}{103.58}
\DefMacro{ds-bf-small-test-avg-source-token-count}{40.54}
\DefMacro{ds-bf-small-test-avg-target-token-count}{34.05}
\DefMacro{ds-bf-small-test-examples-count}{5,831}
\DefMacro{ds-bf-small-train-avg-context-token-count}{77.11}
\DefMacro{ds-bf-small-train-avg-edit-token-count}{52.48}
\DefMacro{ds-bf-small-train-avg-lev-distance}{103.93}
\DefMacro{ds-bf-small-train-avg-source-token-count}{40.57}
\DefMacro{ds-bf-small-train-avg-target-token-count}{33.81}
\DefMacro{ds-bf-small-train-examples-count}{46,628}
\DefMacro{ds-bf-small-valid-avg-context-token-count}{76.66}
\DefMacro{ds-bf-small-valid-avg-edit-token-count}{52.40}
\DefMacro{ds-bf-small-valid-avg-lev-distance}{104.53}
\DefMacro{ds-bf-small-valid-avg-source-token-count}{40.45}
\DefMacro{ds-bf-small-valid-avg-target-token-count}{33.67}
\DefMacro{ds-bf-small-valid-examples-count}{5,828}

\DefMacro{ds-comment-update-test-avg-context-token-count}{175.36}
\DefMacro{ds-comment-update-test-avg-edit-token-count}{22.62}
\DefMacro{ds-comment-update-test-avg-lev-distance}{109.80}
\DefMacro{ds-comment-update-test-avg-source-token-count}{12.47}
\DefMacro{ds-comment-update-test-avg-target-token-count}{13.22}
\DefMacro{ds-comment-update-test-examples-count}{150}
\DefMacro{ds-comment-update-train-avg-context-token-count}{169.19}
\DefMacro{ds-comment-update-train-avg-edit-token-count}{24.36}
\DefMacro{ds-comment-update-train-avg-lev-distance}{112.81}
\DefMacro{ds-comment-update-train-avg-source-token-count}{13.76}
\DefMacro{ds-comment-update-train-avg-target-token-count}{14.52}
\DefMacro{ds-comment-update-train-examples-count}{16,494}
\DefMacro{ds-comment-update-valid-avg-context-token-count}{162.91}
\DefMacro{ds-comment-update-valid-avg-edit-token-count}{23.06}
\DefMacro{ds-comment-update-valid-avg-lev-distance}{108.09}
\DefMacro{ds-comment-update-valid-avg-source-token-count}{12.91}
\DefMacro{ds-comment-update-valid-avg-target-token-count}{13.60}
\DefMacro{ds-comment-update-valid-examples-count}{1,878}

\DefMacro{ds-code-review-test-avg-context-token-count}{24.31}
\DefMacro{ds-code-review-test-avg-edit-token-count}{86.28}
\DefMacro{ds-code-review-test-avg-lev-distance}{118.99}
\DefMacro{ds-code-review-test-avg-source-token-count}{73.39}
\DefMacro{ds-code-review-test-avg-target-token-count}{64.90}
\DefMacro{ds-code-review-test-examples-count}{1,718}
\DefMacro{ds-code-review-train-avg-context-token-count}{24.65}
\DefMacro{ds-code-review-train-avg-edit-token-count}{86.44}
\DefMacro{ds-code-review-train-avg-lev-distance}{119.58}
\DefMacro{ds-code-review-train-avg-source-token-count}{73.47}
\DefMacro{ds-code-review-train-avg-target-token-count}{64.87}
\DefMacro{ds-code-review-train-examples-count}{13,753}
\DefMacro{ds-code-review-valid-avg-context-token-count}{24.33}
\DefMacro{ds-code-review-valid-avg-edit-token-count}{84.68}
\DefMacro{ds-code-review-valid-avg-lev-distance}{119.48}
\DefMacro{ds-code-review-valid-avg-source-token-count}{71.87}
\DefMacro{ds-code-review-valid-avg-target-token-count}{63.34}
\DefMacro{ds-code-review-valid-examples-count}{1,719}

\DefMacro{ds-comment-update-full-test-avg-context-token-count}{168.83}
\DefMacro{ds-comment-update-full-test-avg-edit-token-count}{23.55}
\DefMacro{ds-comment-update-full-test-avg-lev-distance}{109.63}
\DefMacro{ds-comment-update-full-test-avg-source-token-count}{13.32}
\DefMacro{ds-comment-update-full-test-avg-target-token-count}{14.02}
\DefMacro{ds-comment-update-full-test-examples-count}{1,971}
\DefMacro{ds-comment-update-full-train-avg-context-token-count}{112.33}
\DefMacro{ds-comment-update-full-train-avg-edit-token-count}{31.57}
\DefMacro{ds-comment-update-full-train-avg-lev-distance}{124.09}
\DefMacro{ds-comment-update-full-train-avg-source-token-count}{13.76}
\DefMacro{ds-comment-update-full-train-avg-target-token-count}{16.07}
\DefMacro{ds-comment-update-full-train-examples-count}{16,494}
\DefMacro{ds-comment-update-full-valid-avg-context-token-count}{108.36}
\DefMacro{ds-comment-update-full-valid-avg-edit-token-count}{29.97}
\DefMacro{ds-comment-update-full-valid-avg-lev-distance}{118.68}
\DefMacro{ds-comment-update-full-valid-avg-source-token-count}{12.91}
\DefMacro{ds-comment-update-full-valid-avg-target-token-count}{15.11}
\DefMacro{ds-comment-update-full-valid-examples-count}{1,878}

\DefMacro{results-bf-small-plbart-Acc-top-1}{31.09}
\DefMacro{results-bf-small-plbart-BLEU}{66.93}
\DefMacro{results-bf-small-plbart-COPY-PCT}{6.48}
\DefMacro{results-bf-small-plbart-CodeBLEU}{73.06}
\DefMacro{results-bf-small-plbart-GLEU}{58.84}
\DefMacro{results-bf-small-plbart-METEOR}{-1.00}
\DefMacro{results-bf-small-plbart-SARI}{51.49}
\DefMacro{results-bf-small-CodeT5-Acc-top-1}{34.81}
\DefMacro{results-bf-small-CodeT5-BLEU}{68.68}
\DefMacro{results-bf-small-CodeT5-COPY-PCT}{7.97}
\DefMacro{results-bf-small-CodeT5-CodeBLEU}{75.13}
\DefMacro{results-bf-small-CodeT5-GLEU}{60.73}
\DefMacro{results-bf-small-CodeT5-METEOR}{-1.00}
\DefMacro{results-bf-small-CodeT5-SARI}{53.05}
\DefMacro{results-bf-small-CoditT5-ft-Acc-top-1}{36.37}
\DefMacro{results-bf-small-CoditT5-ft-BLEU}{68.19}
\DefMacro{results-bf-small-CoditT5-ft-CodeBLEU}{73.18}
\DefMacro{results-bf-small-CoditT5-ft-GLEU}{61.21}
\DefMacro{results-bf-small-CoditT5-ft-METEOR}{-1.00}
\DefMacro{results-bf-small-CoditT5-ft-SARI}{54.31}
\DefMacro{results-bf-small-CoditT5-Acc-top-1}{37.52}
\DefMacro{results-bf-small-CoditT5-BLEU}{67.81}
\DefMacro{results-bf-small-CoditT5-COPY-PCT}{0.55}
\DefMacro{results-bf-small-CoditT5-CodeBLEU}{70.86}
\DefMacro{results-bf-small-CoditT5-GLEU}{61.11}
\DefMacro{results-bf-small-CoditT5-METEOR}{-1.00}
\DefMacro{results-bf-small-CoditT5-SARI}{53.85}
\DefMacro{results-bf-small-CoditT5-CodeT5-rerank-Acc-top-1}{40.22}
\DefMacro{results-bf-small-CoditT5-CodeT5-rerank-BLEU}{70.68}
\DefMacro{results-bf-small-CoditT5-CodeT5-rerank-COPY-PCT}{1.85}
\DefMacro{results-bf-small-CoditT5-CodeT5-rerank-CodeBLEU}{75.44}
\DefMacro{results-bf-small-CoditT5-CodeT5-rerank-GLEU}{63.48}
\DefMacro{results-bf-small-CoditT5-CodeT5-rerank-METEOR}{-1.00}
\DefMacro{results-bf-small-CoditT5-CodeT5-rerank-SARI}{55.59}
\DefMacro{results-bf-small-CodeT5-CoditT5-rerank-Acc-top-1}{39.56}
\DefMacro{results-bf-small-CodeT5-CoditT5-rerank-BLEU}{69.90}
\DefMacro{results-bf-small-CodeT5-CoditT5-rerank-COPY-PCT}{0.00}
\DefMacro{results-bf-small-CodeT5-CoditT5-rerank-CodeBLEU}{75.99}
\DefMacro{results-bf-small-CodeT5-CoditT5-rerank-GLEU}{62.84}
\DefMacro{results-bf-small-CodeT5-CoditT5-rerank-METEOR}{-1.00}
\DefMacro{results-bf-small-CodeT5-CoditT5-rerank-SARI}{56.02}

\DefMacro{results-bf-medium-plbart-Acc-top-1}{24.18}
\DefMacro{results-bf-medium-plbart-BLEU}{74.80}
\DefMacro{results-bf-medium-plbart-COPY-PCT}{10.92}
\DefMacro{results-bf-medium-plbart-CodeBLEU}{82.99}
\DefMacro{results-bf-medium-plbart-GLEU}{68.32}
\DefMacro{results-bf-medium-plbart-METEOR}{-1.00}
\DefMacro{results-bf-medium-plbart-SARI}{48.81}
\DefMacro{results-bf-medium-CodeT5-Acc-top-1}{26.66}
\DefMacro{results-bf-medium-CodeT5-BLEU}{76.26}
\DefMacro{results-bf-medium-CodeT5-COPY-PCT}{10.08}
\DefMacro{results-bf-medium-CodeT5-CodeBLEU}{83.31}
\DefMacro{results-bf-medium-CodeT5-GLEU}{69.99}
\DefMacro{results-bf-medium-CodeT5-METEOR}{-1.00}
\DefMacro{results-bf-medium-CodeT5-SARI}{50.71}
\DefMacro{results-bf-medium-CoditT5-ft-Acc-top-1}{29.28}
\DefMacro{results-bf-medium-CoditT5-ft-BLEU}{76.58}
\DefMacro{results-bf-medium-CoditT5-ft-CodeBLEU}{83.20}
\DefMacro{results-bf-medium-CoditT5-ft-GLEU}{70.48}
\DefMacro{results-bf-medium-CoditT5-ft-METEOR}{-1.00}
\DefMacro{results-bf-medium-CoditT5-ft-SARI}{51.81}
\DefMacro{results-bf-medium-CoditT5-Acc-top-1}{29.96}
\DefMacro{results-bf-medium-CoditT5-BLEU}{75.74}
\DefMacro{results-bf-medium-CoditT5-COPY-PCT}{0.78}
\DefMacro{results-bf-medium-CoditT5-CodeBLEU}{81.72}
\DefMacro{results-bf-medium-CoditT5-GLEU}{69.89}
\DefMacro{results-bf-medium-CoditT5-METEOR}{-1.00}
\DefMacro{results-bf-medium-CoditT5-SARI}{51.52}
\DefMacro{results-bf-medium-CoditT5-CodeT5-rerank-Acc-top-1}{32.06}
\DefMacro{results-bf-medium-CoditT5-CodeT5-rerank-BLEU}{77.80}
\DefMacro{results-bf-medium-CoditT5-CodeT5-rerank-COPY-PCT}{4.07}
\DefMacro{results-bf-medium-CoditT5-CodeT5-rerank-CodeBLEU}{83.63}
\DefMacro{results-bf-medium-CoditT5-CodeT5-rerank-GLEU}{71.93}
\DefMacro{results-bf-medium-CoditT5-CodeT5-rerank-METEOR}{-1.00}
\DefMacro{results-bf-medium-CoditT5-CodeT5-rerank-SARI}{52.81}
\DefMacro{results-bf-medium-CodeT5-CoditT5-rerank-Acc-top-1}{32.24}
\DefMacro{results-bf-medium-CodeT5-CoditT5-rerank-BLEU}{77.39}
\DefMacro{results-bf-medium-CodeT5-CoditT5-rerank-COPY-PCT}{0.00}
\DefMacro{results-bf-medium-CodeT5-CoditT5-rerank-CodeBLEU}{83.83}
\DefMacro{results-bf-medium-CodeT5-CoditT5-rerank-GLEU}{71.47}
\DefMacro{results-bf-medium-CodeT5-CoditT5-rerank-METEOR}{-1.00}
\DefMacro{results-bf-medium-CodeT5-CoditT5-rerank-SARI}{53.56}

\DefMacro{results-comment-update-EditModel-Acc-top-1}{33.33}
\DefMacro{results-comment-update-EditModel-BLEU}{56.55}
\DefMacro{results-comment-update-EditModel-CodeBLEU}{-1.00}
\DefMacro{results-comment-update-EditModel-GLEU}{51.88}
\DefMacro{results-comment-update-EditModel-METEOR}{52.26}
\DefMacro{results-comment-update-EditModel-SARI}{56.23}
\DefMacro{results-comment-update-plbart-Acc-top-1}{35.33}
\DefMacro{results-comment-update-plbart-BLEU}{62.04}
\DefMacro{results-comment-update-plbart-COPY-PCT}{21.33}
\DefMacro{results-comment-update-plbart-CodeBLEU}{-1.00}
\DefMacro{results-comment-update-plbart-GLEU}{54.75}
\DefMacro{results-comment-update-plbart-METEOR}{56.79}
\DefMacro{results-comment-update-plbart-SARI}{52.83}
\DefMacro{results-comment-update-CodeT5-Acc-top-1}{38.00}
\DefMacro{results-comment-update-CodeT5-BLEU}{65.20}
\DefMacro{results-comment-update-CodeT5-COPY-PCT}{16.67}
\DefMacro{results-comment-update-CodeT5-CodeBLEU}{-1.00}
\DefMacro{results-comment-update-CodeT5-GLEU}{58.84}
\DefMacro{results-comment-update-CodeT5-METEOR}{59.63}
\DefMacro{results-comment-update-CodeT5-SARI}{58.80}
\DefMacro{results-comment-update-CoditT5-ft-Acc-top-1}{40.00}
\DefMacro{results-comment-update-CoditT5-ft-BLEU}{62.97}
\DefMacro{results-comment-update-CoditT5-ft-COPY-PCT}{2.00}
\DefMacro{results-comment-update-CoditT5-ft-CodeBLEU}{-1.00}
\DefMacro{results-comment-update-CoditT5-ft-GLEU}{58.72}
\DefMacro{results-comment-update-CoditT5-ft-METEOR}{59.08}
\DefMacro{results-comment-update-CoditT5-ft-SARI}{61.11}
\DefMacro{results-comment-update-CoditT5-Acc-top-1}{43.33}
\DefMacro{results-comment-update-CoditT5-BLEU}{64.56}
\DefMacro{results-comment-update-CoditT5-COPY-PCT}{2.67}
\DefMacro{results-comment-update-CoditT5-CodeBLEU}{-1.00}
\DefMacro{results-comment-update-CoditT5-GLEU}{59.53}
\DefMacro{results-comment-update-CoditT5-METEOR}{60.75}
\DefMacro{results-comment-update-CoditT5-SARI}{61.41}
\DefMacro{results-comment-update-CoditT5-CodeT5-rerank-Acc-top-1}{45.33}
\DefMacro{results-comment-update-CoditT5-CodeT5-rerank-BLEU}{66.80}
\DefMacro{results-comment-update-CoditT5-CodeT5-rerank-COPY-PCT}{11.33}
\DefMacro{results-comment-update-CoditT5-CodeT5-rerank-CodeBLEU}{-1.00}
\DefMacro{results-comment-update-CoditT5-CodeT5-rerank-GLEU}{61.60}
\DefMacro{results-comment-update-CoditT5-CodeT5-rerank-METEOR}{63.33}
\DefMacro{results-comment-update-CoditT5-CodeT5-rerank-SARI}{61.48}
\DefMacro{results-comment-update-CodeT5-CoditT5-rerank-Acc-top-1}{44.00}
\DefMacro{results-comment-update-CodeT5-CoditT5-rerank-BLEU}{65.58}
\DefMacro{results-comment-update-CodeT5-CoditT5-rerank-COPY-PCT}{1.33}
\DefMacro{results-comment-update-CodeT5-CoditT5-rerank-CodeBLEU}{-1.00}
\DefMacro{results-comment-update-CodeT5-CoditT5-rerank-GLEU}{60.48}
\DefMacro{results-comment-update-CodeT5-CoditT5-rerank-METEOR}{62.44}
\DefMacro{results-comment-update-CodeT5-CoditT5-rerank-SARI}{62.57}

\DefMacro{results-comment-update-full-EditModel-Acc-top-1}{24.81}
\DefMacro{results-comment-update-full-EditModel-BLEU}{48.89}
\DefMacro{results-comment-update-full-EditModel-CodeBLEU}{-1.00}
\DefMacro{results-comment-update-full-EditModel-GLEU}{45.69}
\DefMacro{results-comment-update-full-EditModel-METEOR}{44.58}
\DefMacro{results-comment-update-full-EditModel-SARI}{47.93}
\DefMacro{results-comment-update-full-plbart-Acc-top-1}{22.98}
\DefMacro{results-comment-update-full-plbart-BLEU}{55.42}
\DefMacro{results-comment-update-full-plbart-CodeBLEU}{-1.00}
\DefMacro{results-comment-update-full-plbart-GLEU}{47.83}
\DefMacro{results-comment-update-full-plbart-METEOR}{49.12}
\DefMacro{results-comment-update-full-plbart-SARI}{43.40}
\DefMacro{results-comment-update-full-plbart-COPY-PCT}{34.25}
\DefMacro{results-comment-update-full-CodeT5-Acc-top-1}{28.56}
\DefMacro{results-comment-update-full-CodeT5-BLEU}{58.37}
\DefMacro{results-comment-update-full-CodeT5-COPY-PCT}{25.47}
\DefMacro{results-comment-update-full-CodeT5-CodeBLEU}{-1.00}
\DefMacro{results-comment-update-full-CodeT5-GLEU}{51.90}
\DefMacro{results-comment-update-full-CodeT5-METEOR}{53.13}
\DefMacro{results-comment-update-full-CodeT5-SARI}{49.23}
\DefMacro{results-comment-update-full-CoditT5-ft-Acc-top-1}{29.83}
\DefMacro{results-comment-update-full-CoditT5-ft-BLEU}{54.83}
\DefMacro{results-comment-update-full-CoditT5-ft-COPY-PCT}{2.79}
\DefMacro{results-comment-update-full-CoditT5-ft-CodeBLEU}{-1.00}
\DefMacro{results-comment-update-full-CoditT5-ft-GLEU}{50.67}
\DefMacro{results-comment-update-full-CoditT5-ft-METEOR}{50.71}
\DefMacro{results-comment-update-full-CoditT5-ft-SARI}{52.01}
\DefMacro{results-comment-update-full-CoditT5-Acc-top-1}{29.38}
\DefMacro{results-comment-update-full-CoditT5-BLEU}{55.30}
\DefMacro{results-comment-update-full-CoditT5-COPY-PCT}{5.73}
\DefMacro{results-comment-update-full-CoditT5-CodeBLEU}{-1.00}
\DefMacro{results-comment-update-full-CoditT5-GLEU}{50.62}
\DefMacro{results-comment-update-full-CoditT5-METEOR}{51.14}
\DefMacro{results-comment-update-full-CoditT5-SARI}{51.39}
\DefMacro{results-comment-update-full-CoditT5-CodeT5-rerank-Acc-top-1}{30.14}
\DefMacro{results-comment-update-full-CoditT5-CodeT5-rerank-BLEU}{58.72}
\DefMacro{results-comment-update-full-CoditT5-CodeT5-rerank-COPY-PCT}{19.28}
\DefMacro{results-comment-update-full-CoditT5-CodeT5-rerank-CodeBLEU}{-1.00}
\DefMacro{results-comment-update-full-CoditT5-CodeT5-rerank-GLEU}{52.81}
\DefMacro{results-comment-update-full-CoditT5-CodeT5-rerank-METEOR}{53.60}
\DefMacro{results-comment-update-full-CoditT5-CodeT5-rerank-SARI}{50.47}
\DefMacro{results-comment-update-full-CodeT5-CoditT5-rerank-Acc-top-1}{27.80}
\DefMacro{results-comment-update-full-CodeT5-CoditT5-rerank-BLEU}{55.54}
\DefMacro{results-comment-update-full-CodeT5-CoditT5-rerank-COPY-PCT}{2.18}
\DefMacro{results-comment-update-full-CodeT5-CoditT5-rerank-CodeBLEU}{-1.00}
\DefMacro{results-comment-update-full-CodeT5-CoditT5-rerank-GLEU}{50.02}
\DefMacro{results-comment-update-full-CodeT5-CoditT5-rerank-METEOR}{51.44}
\DefMacro{results-comment-update-full-CodeT5-CoditT5-rerank-SARI}{52.24}

\DefMacro{results-code-review-plbart-Acc-top-1}{26.78}
\DefMacro{results-code-review-plbart-BLEU}{79.38}
\DefMacro{results-code-review-plbart-COPY-PCT}{22.24}
\DefMacro{results-code-review-plbart-CodeBLEU}{83.22}
\DefMacro{results-code-review-plbart-GLEU}{71.64}
\DefMacro{results-code-review-plbart-METEOR}{-1.00}
\DefMacro{results-code-review-plbart-SARI}{51.85}
\DefMacro{results-code-review-CodeT5-Acc-top-1}{34.98}
\DefMacro{results-code-review-CodeT5-BLEU}{83.20}
\DefMacro{results-code-review-CodeT5-COPY-PCT}{29.28}
\DefMacro{results-code-review-CodeT5-CodeBLEU}{87.53}
\DefMacro{results-code-review-CodeT5-GLEU}{75.74}
\DefMacro{results-code-review-CodeT5-METEOR}{-1.00}
\DefMacro{results-code-review-CodeT5-SARI}{55.90}
\DefMacro{results-code-review-CoditT5-ft-Acc-top-1}{36.38}
\DefMacro{results-code-review-CoditT5-ft-BLEU}{80.06}
\DefMacro{results-code-review-CoditT5-ft-CodeBLEU}{83.75}
\DefMacro{results-code-review-CoditT5-ft-GLEU}{73.70}
\DefMacro{results-code-review-CoditT5-ft-METEOR}{-1.00}
\DefMacro{results-code-review-CoditT5-ft-SARI}{57.99}
\DefMacro{results-code-review-CoditT5-Acc-top-1}{37.19}
\DefMacro{results-code-review-CoditT5-BLEU}{80.50}
\DefMacro{results-code-review-CoditT5-COPY-PCT}{1.28}
\DefMacro{results-code-review-CoditT5-CodeBLEU}{84.19}
\DefMacro{results-code-review-CoditT5-GLEU}{74.18}
\DefMacro{results-code-review-CoditT5-METEOR}{-1.00}
\DefMacro{results-code-review-CoditT5-SARI}{58.66}
\DefMacro{results-code-review-CoditT5-CodeT5-rerank-Acc-top-1}{40.98}
\DefMacro{results-code-review-CoditT5-CodeT5-rerank-BLEU}{84.12}
\DefMacro{results-code-review-CoditT5-CodeT5-rerank-COPY-PCT}{17.40}
\DefMacro{results-code-review-CoditT5-CodeT5-rerank-CodeBLEU}{87.85}
\DefMacro{results-code-review-CoditT5-CodeT5-rerank-GLEU}{77.48}
\DefMacro{results-code-review-CoditT5-CodeT5-rerank-METEOR}{-1.00}
\DefMacro{results-code-review-CoditT5-CodeT5-rerank-SARI}{59.13}
\DefMacro{results-code-review-CodeT5-CoditT5-rerank-Acc-top-1}{43.42}
\DefMacro{results-code-review-CodeT5-CoditT5-rerank-BLEU}{83.92}
\DefMacro{results-code-review-CodeT5-CoditT5-rerank-COPY-PCT}{0.00}
\DefMacro{results-code-review-CodeT5-CoditT5-rerank-CodeBLEU}{87.38}
\DefMacro{results-code-review-CodeT5-CoditT5-rerank-GLEU}{77.73}
\DefMacro{results-code-review-CodeT5-CoditT5-rerank-METEOR}{-1.00}
\DefMacro{results-code-review-CodeT5-CoditT5-rerank-SARI}{61.94}

\DefMacro{valid-bf-small-MOEDIT-PLBART-avg-self-bleu}{70.39}
\DefMacro{valid-bf-small-MOEDIT-PLBART-diff-nMatch-avg-self-bleu}{55.91}
\DefMacro{valid-bf-small-MOEDIT-PLBART-diff-oneMatch-avg-self-bleu}{52.42}
\DefMacro{valid-bf-small-MOEDIT-PLBART-diff-output-avg-self-bleu}{54.92}
\DefMacro{valid-bf-small-MOEDIT-PLBART-optimal-avg-bleu}{74.47}
\DefMacro{valid-bf-small-MOEDIT-PLBART-same-output-avg-bleu}{80.02}
\DefMacro{valid-bf-small-MOEDIT-PLBART-same-output-count}{2,001}
\DefMacro{valid-bf-small-MOEDIT-PLBART-same-output-nMatch-avg-bleu}{48.94}
\DefMacro{valid-bf-small-MOEDIT-PLBART-same-output-xMatch}{1,218}
\DefMacro{valid-bf-small-MOEDIT-avg-bleu}{65.45}

\DefMacro{valid-bf-small-MOEDIT-nMatch-PLBART-nMatch-count}{3,519}
\DefMacro{valid-bf-small-MOEDIT-nMatch-PLBART-xMatch-count}{492}
\DefMacro{valid-bf-small-MOEDIT-xMatch-PLBART-nMatch-count}{599}
\DefMacro{valid-bf-small-MOEDIT-xMatch-PLBART-xMatch-count}{1,218}

\DefMacro{valid-bf-medium-MOEDIT-PLBART-avg-self-bleu}{77.32}
\DefMacro{valid-bf-medium-MOEDIT-PLBART-diff-nMatch-avg-self-bleu}{69.72}
\DefMacro{valid-bf-medium-MOEDIT-PLBART-diff-oneMatch-avg-self-bleu}{63.90}
\DefMacro{valid-bf-medium-MOEDIT-PLBART-diff-output-avg-self-bleu}{68.48}
\DefMacro{valid-bf-medium-MOEDIT-PLBART-optimal-avg-bleu}{80.30}
\DefMacro{valid-bf-medium-MOEDIT-PLBART-same-output-avg-bleu}{85.65}
\DefMacro{valid-bf-medium-MOEDIT-PLBART-same-output-count}{1,835}
\DefMacro{valid-bf-medium-MOEDIT-PLBART-same-output-nMatch-avg-bleu}{65.59}
\DefMacro{valid-bf-medium-MOEDIT-PLBART-same-output-xMatch}{1,070}
\DefMacro{valid-bf-medium-MOEDIT-avg-bleu}{73.36}

\DefMacro{valid-bf-medium-MOEDIT-nMatch-PLBART-nMatch-count}{4,472}
\DefMacro{valid-bf-medium-MOEDIT-nMatch-PLBART-xMatch-count}{514}
\DefMacro{valid-bf-medium-MOEDIT-xMatch-PLBART-nMatch-count}{486}
\DefMacro{valid-bf-medium-MOEDIT-xMatch-PLBART-xMatch-count}{1,070}

\DefMacro{valid-comment-update-MOEDIT-PLBART-avg-self-bleu}{69.37}
\DefMacro{valid-comment-update-MOEDIT-PLBART-diff-nMatch-avg-self-bleu}{53.34}
\DefMacro{valid-comment-update-MOEDIT-PLBART-diff-oneMatch-avg-self-bleu}{58.75}
\DefMacro{valid-comment-update-MOEDIT-PLBART-diff-output-avg-self-bleu}{54.13}
\DefMacro{valid-comment-update-MOEDIT-PLBART-optimal-avg-bleu}{61.54}
\DefMacro{valid-comment-update-MOEDIT-PLBART-same-output-avg-bleu}{80.16}
\DefMacro{valid-comment-update-MOEDIT-PLBART-same-output-count}{624}
\DefMacro{valid-comment-update-MOEDIT-PLBART-same-output-nMatch-avg-bleu}{44.23}
\DefMacro{valid-comment-update-MOEDIT-PLBART-same-output-xMatch}{402}
\DefMacro{valid-comment-update-MOEDIT-avg-bleu}{55.48}

\DefMacro{valid-comment-update-MOEDIT-nMatch-PLBART-nMatch-count}{1,292}
\DefMacro{valid-comment-update-MOEDIT-nMatch-PLBART-xMatch-count}{47}
\DefMacro{valid-comment-update-MOEDIT-xMatch-PLBART-nMatch-count}{137}
\DefMacro{valid-comment-update-MOEDIT-xMatch-PLBART-xMatch-count}{402}

\DefMacro{valid-code-review-MOEDIT-PLBART-avg-self-bleu}{82.29}
\DefMacro{valid-code-review-MOEDIT-PLBART-diff-nMatch-avg-self-bleu}{75.60}
\DefMacro{valid-code-review-MOEDIT-PLBART-diff-oneMatch-avg-self-bleu}{72.14}
\DefMacro{valid-code-review-MOEDIT-PLBART-diff-output-avg-self-bleu}{74.73}
\DefMacro{valid-code-review-MOEDIT-PLBART-optimal-avg-bleu}{84.52}
\DefMacro{valid-code-review-MOEDIT-PLBART-same-output-avg-bleu}{89.93}
\DefMacro{valid-code-review-MOEDIT-PLBART-same-output-count}{514}
\DefMacro{valid-code-review-MOEDIT-PLBART-same-output-nMatch-avg-bleu}{75.70}
\DefMacro{valid-code-review-MOEDIT-PLBART-same-output-xMatch}{301}
\DefMacro{valid-code-review-MOEDIT-avg-bleu}{76.90}

\DefMacro{valid-code-review-MOEDIT-nMatch-PLBART-nMatch-count}{1,117}
\DefMacro{valid-code-review-MOEDIT-nMatch-PLBART-xMatch-count}{181}
\DefMacro{valid-code-review-MOEDIT-xMatch-PLBART-nMatch-count}{120}
\DefMacro{valid-code-review-MOEDIT-xMatch-PLBART-xMatch-count}{301}

\DefMacro{test-bf-small-CoditT5-xMatch-pct}{24.40}
\DefMacro{test-bf-small-CodeT5-xMatch-pct}{18.33}
\DefMacro{test-bf-small-Both-xMatch-pct}{57.27}
\DefMacro{test-bf-medium-CoditT5-xMatch-pct}{26.81}
\DefMacro{test-bf-medium-CodeT5-xMatch-pct}{21.73}
\DefMacro{test-bf-medium-Both-xMatch-pct}{51.46}
\DefMacro{test-comment-update-CoditT5-xMatch-pct}{17.14}
\DefMacro{test-comment-update-CodeT5-xMatch-pct}{8.57}
\DefMacro{test-comment-update-Both-xMatch-pct}{74.29}
\DefMacro{test-code-review-CoditT5-xMatch-pct}{22.12}
\DefMacro{test-code-review-CodeT5-xMatch-pct}{19.15}
\DefMacro{test-code-review-Both-xMatch-pct}{58.73}

\DefMacro{CoditT5-nMatch-CodeT5-nMatch-bf-small}{0.54}
\DefMacro{CoditT5-nMatch-CodeT5-xMatch-bf-small}{0.08}
\DefMacro{CoditT5-xMatch-CodeT5-nMatch-bf-small}{0.11}
\DefMacro{CoditT5-xMatch-CodeT5-xMatch-bf-small}{0.26}
\DefMacro{CoditT5-nMatch-CodeT5-nMatch-bf-medium}{0.62}
\DefMacro{CoditT5-nMatch-CodeT5-xMatch-bf-medium}{0.08}
\DefMacro{CoditT5-xMatch-CodeT5-nMatch-bf-medium}{0.10}
\DefMacro{CoditT5-xMatch-CodeT5-xMatch-bf-medium}{0.20}
\DefMacro{CoditT5-nMatch-CodeT5-nMatch-comment-update}{0.53}
\DefMacro{CoditT5-nMatch-CodeT5-xMatch-comment-update}{0.04}
\DefMacro{CoditT5-xMatch-CodeT5-nMatch-comment-update}{0.08}
\DefMacro{CoditT5-xMatch-CodeT5-xMatch-comment-update}{0.35}
\DefMacro{CoditT5-nMatch-CodeT5-nMatch-code-review}{0.55}
\DefMacro{CoditT5-nMatch-CodeT5-xMatch-code-review}{0.09}
\DefMacro{CoditT5-xMatch-CodeT5-nMatch-code-review}{0.10}
\DefMacro{CoditT5-xMatch-CodeT5-xMatch-code-review}{0.26}

\DefMacro{metrics-delete-prob-PL}{0.49}
\DefMacro{metrics-insert-prob-PL}{0.21}
\DefMacro{metrics-replace-prob-PL}{0.30}
\DefMacro{metrics-mean-span-length-PL}{6.50}
\DefMacro{metrics-mean-span-count-PL}{1.90}

\DefMacro{metrics-delete-prob-NL}{0.07}
\DefMacro{metrics-insert-prob-NL}{0.11}
\DefMacro{metrics-replace-prob-NL}{0.82}
\DefMacro{metrics-mean-span-length-NL}{3.00}
\DefMacro{metrics-mean-span-count-NL}{1.40}

\title[\ShortTitle]{\Title}

\author{Jiyang Zhang}
\affiliation{
  \institution{The University of Texas at Austin}
  \city{Austin}
  \state{TX}
  \country{USA}
}
\email{jiyang.zhang@utexas.edu}

\author{Sheena Panthaplackel}
\affiliation{
  \institution{The University of Texas at Austin}
  \city{Austin}
  \state{TX}
  \country{USA}
}
\email{spantha@cs.utexas.edu}

\author{Pengyu Nie}
\affiliation{
  \institution{The University of Texas at Austin}
  \city{Austin}
  \state{TX}
  \country{USA}
}
\email{pynie@utexas.edu}

\author{Junyi Jessy Li}
\affiliation{
  \institution{The University of Texas at Austin}
  \city{Austin}
  \state{TX}
  \country{USA}
}
\email{jessy@austin.utexas.edu}

\author{Milos Gligoric}
\affiliation{
  \institution{The University of Texas at Austin}
  \city{Austin}
  \state{TX}
  \country{USA}
}
\email{gligoric@utexas.edu}

\copyrightyear{2022}
\acmYear{2022}
\setcopyright{acmlicensed}\acmConference[ASE '22]{37th IEEE/ACM International Conference on Automated Software Engineering}{October 10--14, 2022}{Rochester, MI, USA}
\acmBooktitle{37th IEEE/ACM International Conference on Automated Software Engineering (ASE '22), October 10--14, 2022, Rochester, MI, USA}
\acmPrice{15.00}
\acmDOI{10.1145/3551349.3556955}
\acmISBN{978-1-4503-9475-8/22/10}

\begin{document}
\begin{abstract}
    \Pretrained language models have been shown to be effective in many software-related
    generation tasks; however, they are not well-suited for editing tasks as they are not
    designed to reason about edits. To address this, we propose a novel \pretraining objective
    which explicitly models edits and use it to build \MOEDIT, a large language model for software-related editing tasks that is \pretrained on large amounts of source code and natural language comments. We \finetune it on various downstream editing tasks, including \cupdate, \bugfix, and \codereview. By outperforming standard generation-based models, we demonstrate the generalizability of our approach and its suitability for editing tasks. We also show how a standard generation model and our edit-based model can complement one another through simple reranking strategies, with which we achieve state-of-the-art performance for the three downstream editing tasks.
\end{abstract}



\maketitle

\section{Introduction}

Large language models \pretrained on massive amounts of data have led
to remarkable progress in recent years, with models like
BART~\cite{lewis-etal-2020-bart}, GPT~\cite{radford2019language,brown2020language}, and T5~\cite{raffel2020exploring} yielding
huge improvements for a vast number of text generation tasks. Inspired
by this, a new research initiative has emerged around building large
models that are \pretrained on source code and technical text to
address software-related tasks. This includes models like
PLBART~\cite{ahmad-etal-2021-unified},
CodeGPT-2~\cite{lu2021codexglue}, and
CodeT5~\cite{wang2021codet5}. While these models demonstrate
impressive performance on generation tasks like code summarization,
code generation, and code translation,
it is unclear if they are well-suited for the \textit{editing} nature of many software-related tasks.
For instance, \bugfix~\cite{TufanoETAL19Empirical} entails editing source
code to resolve bugs, \codereview~\cite{Tufano21Towards} requires editing source
code to incorporate feedback from review comments, and
\cupdate~\cite{PanthaplackelETAL20Learning, LiuETAL21Just, LinETAL21Automated, GaoETAL21Automating} pertains to updating
outdated natural language comments to reflect code changes.

In principle, such editing tasks can be framed as standard generation
tasks in which an input sequence (e.g., \emph{buggy} code snippet) is
completely re-written to form the output sequence (e.g., \emph{fixed}
code snippet).
In this way, existing \pretrained conditional generation models can be fine-tuned to autoregressively generate a sequence from scratch. However, this can be problematic in practice~\cite{PanthaplackelETAL20Learning}.
When applying large generation models like
PLBART and CodeT5 to these tasks, we find that they can
generate
output which merely copies the input without performing any edits (up to \HighestCopyRate) or
even deviates substantially from the input, introducing irrelevant
changes.  
We provide an example of \codereview in
Figure~\ref{fig:intro-example:1}, where a reviewer prescribes edits that need to be made
to a given code snippet: ``\introReviewerComment''. Using the code snippet and this comment, PLBART
generates an output sequence which copies the original code, without
applying any edits. While the output is valid and a likely sequence
according to PLBART's language model, it makes no edits based on
the reviewer's comments.

We attribute these weaknesses to the fact that such models rely on \pretraining objectives designed for generating code (or software-related natural language) in sequence by exploiting patterns with respect to preceding tokens. Therefore, a model
has to
learn to
\textit{implicitly} perform edits by generating tokens one by one in accordance with the underlying probability that it has learned for which tokens belong alongside one another, rather than being aware of where information should be retained or modified.

Intuitively, edit-based generation requires a different approach that
more frequently refers back to the input sequence, and can often be
characterized by localized operations (e.g., insertion, deletion,
substitution). To guide a model in discerning edit locations in the input sequence and reason about the necessary edit operations, we design
a novel \pretraining objective that
\textit{explicitly} models edits.
Our approach is inspired by content planning in natural language generation where a skeleton of key elements are first generated and used to guide more accurate and precise generation of full text~\cite{reiter1997building,pichotta2016learning,martin2018event,fan2019strategies}.
Specifically, during decoding, a model first generates an \textit{edit plan} that explicitly details the edit
operations. Then, it proceeds to autoregressively generate the target edited sequence, during which it attends to the edit plan. Through this, we effectively encourage the model to learn to better reason about edits and how they should be applied to form the target sequence.
Using this objective, we develop \MOEDIT, a large language model for software-related edit tasks that is \pretrained
on more than \UseMacro{pl-pretrain-number}~million open-source programming language\XComment{ functions} code snippets and \UseMacro{nl-pretrain-number}~million natural language comments from the CodeSearchNet~\cite{HusainETAL19Codesearchnet} training data.

For evaluation, we \finetune \MOEDIT on three downstream tasks: \cupdate, \bugfix,  and \codereview. For each of these tasks, we show that \MOEDIT outperforms state-of-the-art models as well as large \pretrained standard generation-based models. Through this, we demonstrate that our model and the proposed edit-based \pretraining objective generalize across tasks and are better suited for editing tasks in the software domain.

Furthermore, in our evaluation, we find that our edit-based model, \MOEDIT,
can be further improved if combined with a standard generation-based model. We find that the edit-based and standard generation-based models are complementary to one another. Namely, while the edit-based model provides better explicit modeling of concrete edits, a standard generation-based model provides certain advantages in terms of the contextual coherence of the generated target sequence.
To exploit this complementary nature of these models,
we combine the two models through reranking strategies which require no additional training. Our results show that the combined approaches outperform the two models individually by up to \UseMacro{rerank-best-improve}\%.

\noindent
We summarize our main contributions as follows:

\begin{itemize}[topsep=3pt,itemsep=1ex,partopsep=0ex,parsep=0ex,leftmargin=*]
	\item We formulate a novel \pretraining objective that entails first generating a plan consisting of edit operations to be applied to the input sequence followed by the resulting target sequence.
	\item We build and release \MOEDIT, a large language model for software-related editing tasks that is \pretrained on large amounts of source code and natural language with the new \pretraining objective.
	\item Upon task-specific fine-tuning, we show that \MOEDIT achieves improved performance over existing models for three distinct downstream editing tasks (\cupdate, \bugfix and \codereview), demonstrating its effectiveness and generalizability.
	\item We show that by combining our edit-based \MOEDIT model with a standard generation model through simple reranking strategies, we can beat each of the individual models and achieve new state-of-the-art in all three tasks, demonstrating the complementary nature of edit-based and standard generation models.
\end{itemize}

\begin{figure}[t]
\centering
\renewcommand{\arraystretch}{1} 
\begin{tabular}{l}
\toprule
\underline{\small Before Editing} \\
{
\begin{lstlisting}[language=java-pretty, numbers=none]
default List<Pattern> getExcludedResponseHeaderPatterns() {
    return emptyList();
}
\end{lstlisting}}
\\
\underline{\small Reviewer's Comment}\\
{\small \introReviewerComment}\\
\underline{\small PLBART} \\
       {
\begin{lstlisting}[language=java-pretty, numbers=none]
default List<Pattern> getExcludedResponseHeaderPatterns() {
    return emptyList();
}
\end{lstlisting}}\\

\bottomrule
\end{tabular}
\caption{\UseMacro{FCap-example-moedit-better}}\label{fig:intro-example:1}
\end{figure}




\noindent
Our code and data is publicly available at\\ \url{https://github.com/EngineeringSoftware/CoditT5}.

\begin{figure*}[t]
	\centering
	\begin{tikzpicture}[
        node distance = 5cm,
        every node/.style={scale=1},
        roundnode/.style={circle, draw, thin, minimum size=1em},
    ]
    \node [block] at (0,0) (encoder) {Encoder};
    \node [text_box, above = 4ex of encoder, font=\ttfamily, xshift=2em, text width=20em] (input) {\small @param \texttt{[MASK]} List of objects};
    \node [text_box, above = 4ex of input, font=\ttfamily, text width=20em] (original) {\small @param users List of user objects};
    \node [block, right = 5em of encoder] (decoder) {Decoder};
    \node [long_text_box, minimum height=0.6cm, below = 4ex of decoder, font=\ttfamily, xshift=2em, text width=34em] (edits) {\small  <ReplaceOld> \texttt{[MASK]} <ReplaceNew> users <ReplaceEnd> <Insert> user <InsertEnd> };
    \node [text_box, below = 0 of edits, minimum height=1mm, font=\ttfamily, xshift=-2em, text width=20em] (sep) {\small<s>};
    \node [text_box, below = 0 of sep, minimum height=1mm, font=\ttfamily, text width=20em] (output) {\small@param users List of user objects};
    \node[shape=circle, draw=black, minimum size=0.1, inner sep=2pt, fill=white, thick, text=black, right = 0.2mm of edits] (label1) {\scriptsize 1};
    \node[shape=circle, draw=black, minimum size=0.1, inner sep=2pt, fill=white, thick, text=black, right = 0.2mm of output] (label1) {\scriptsize 2};

    \path [line] (original.south -| encoder.north) -- node [right] {noising function} (input.north -| encoder.north);
    \path [line] (input.south -| encoder.north) -- (encoder.north);
    \path [line] (encoder) -- (decoder);
    \path [line] (decoder.south) -- (edits.north -| decoder.south);

\end{tikzpicture}
	\caption{\UseMacro{FCap-pretrain-model} \label{fig:pretrain}}
	\label{fig:pretrain-model}
\end{figure*}
\section{Background}

We first give a high-level overview of the building blocks that are necessary to understand our approach.

\subsection{Generation with Transformer-Based Models}
\paragraph{Conditional Sequence Generation}
Conditional sequence generation entails generating an output sequence
given an input sequence. Many tasks are framed in this manner,
including machine translation (e.g., translating a sentence from
French to English)~\cite{BahdanauETAL15Attention}, text summarization
(e.g., generating a brief summary for a given news
article)~\cite{RushETAL15Summarization}, and code generation (e.g.,
generating a code snippet for a given natural language
specification)~\cite{YinAndNeubig17CodeGen}.

\paragraph{Encoder-Decoder Framework}
In recent years, conditional sequence generation tasks are being
addressed with encoder-decoder models. An encoder-decoder model
consists of two neural components: an encoder and a decoder. The input
sequence is fed into the encoder, which produces learned vector
representations of the tokens in that sequence.  These learned vector
representations are then passed into the decoder, which generates the
output sequence one token at a time.  Specifically, the decoder
predicts the next token by reasoning over the input sequence and the tokens
generated at previous time steps.

\paragraph{Transformers}
Transformers~\cite{VaswaniETAL17Attention} are powerful neural models
that are commonly adopted as the encoder and decoder in the
encoder-decoder framework. These models rely on an \textit{attention}
mechanism to learn representations for tokens by relating them to
other tokens in the sequence. Namely, a transformer-based encoder will
learn representations for each token in the input sequence by
``attending'' to other input tokens. For the decoder, when generating
a token at timestep $t$, it will ``attend'' to the representations of
the output tokens generated from timestep 1 to $t-1$ as well as the
representations of tokens from the input sequence. Transformer models
can become very large with huge numbers of attention heads, encoder
and decoder layers.

\subsection{Large \Pretrained Language Models}

Large \pretrained language models generally refer to the class of
large transformer-based models that are trained on large amounts of
unlabeled data (collected from webpages, news articles, etc.) with
unsupervised training objectives. This includes a vast number of
models like GPT~\cite{radford2019language,brown2020language},
BART~\cite{lewis-etal-2020-bart}, and T5~\cite{raffel2020exploring}.

\paragraph{Denoising Autoencoder Pretraining}
BART and T5 models are \pretrained using denoising autoencoding unsupervised training objectives. Namely, a noising function is first applied to a given input sequence \inp to form \noisyinp. Common noising functions include \textit{Token Masking}: tokens in the input sequence are randomly masked;
\textit{Token Deletion}: random tokens are deleted from the input sequence;
\textit{Token Infilling}: a span of tokens are sampled and replaced with a mask token;
\textit{Sentence Permutation}: sentences in the document are shuffled in a random order.
Then, \noisyinp is fed into a model's encoder, and the encoder's learned representation is passed into the decoder, which generates an output sequence, \out, that is expected to resemble the original input sequence (\inp). In other words, the model is trained to ``denoise'' \noisyinp, using a training objective that minimizes the error between \out and the original input, \inp. 
Through this, the model learns to extract meaning from the input sequence and also generate fluent and coherent output. Therefore, by \pretraining on massive amounts of data, the model develops an understanding of how things in the world relate to one another as a strong language modeling capability.

\paragraph{\Finetuning for Downstream Tasks}
Since large \pretrained language models are trained using unsupervised
training objectives on huge amounts of data, they cannot generally be
directly applied to downstream tasks (e.g., translation,
summarization).  \Finetuning is a common technique to transfer the
knowledge learned during \pretraining to target downstream tasks.
Specifically, the \pretrained model is further trained for the
downstream task on some amount of supervised data.

\subsection{Large \Pretrained Language Models for Software Engineering}
\label{sec:CodeT5}
Inspired by the success of large \pretrained models in Natural Language Processing (NLP), a number of machine learning models \pretrained on source code and technical text have been proposed
for solving various software-related problems.

For instance, inspired by BART, \citet{ahmad-etal-2021-unified}
developed PLBART, which is a large \pretrained language model that can
be \finetuned for a number of code understanding (e.g., code
summarization) and generation (e.g., code translation) tasks.
Similarly, inspired by T5, \citet{wang2021codet5} built a larger model
\CodeTf, which is \pretrained on six programming languages together
with their natural language comments collected from open-source
repositories.  Specially, it is \pretrained to incorporate information
from identifiers in the code.  \CodeTf has shown promising results in
code-related generation tasks such as code summarization, code
generation and code-related understanding tasks such as clone
detection and vulnerability identification.
However, aforementioned models are for generation and they are only
implicitly aware of edit operations if at all.

\section{CoditT5}

\MOEDIT is built upon the encoder-decoder framework with the same
architecture as \CodeTf.  As shown in Figure~\ref{fig:pretrain-model},
the model is \pretrained with our proposed objective: generating the
edit-based output sequence given the corrupted input sequence.  In
this section, we first explain our proposed \pretraining objective
(Section~\ref{sub:pretraining}). We then discuss how we build \MOEDIT
by \pretraining on this objective, including the data used for
\pretraining (Section~\ref{sub:pretrain_data}), and additional details
of the \pretraining setup (Section~\ref{sub:pretrain_setup}).

\subsection{\Pretraining Objective}
\label{sub:pretraining}

We formulate a new \pretraining objective that is designed to encourage
a model to explicitly reason about edits. At a high-level, this
objective falls under the realm of denoising autoencoding in which an
input sequence is first corrupted with noising functions and the model
is trained to \textit{denoise} the corrupted sequence by generating an
output sequence that matches the original input sequence. While existing models like PLBART and \CodeTf \pretrained using this setup perform very well on various generation tasks (e.g., code summarization/generation), we find that they do not generalize well when \finetuned on editing tasks. Namely, they are susceptible to learning to copy the original input sequence instead of actually performing edits, up to \HighestCopyRate of the time (Table~\ref{tab:copy-pct-results}).

We propose the following \textit{edit-based output sequence}
representation (shown in Figure~\ref{fig:pretrain-model}): [Edit Plan] \texttt{<s>} [Target Sequence], where the
model is trained to generate an \textit{edit plan} (\circled{1}) consisting of
explicit edit operations that must be applied to the corrupted sequence to
reconstruct the original input sequence, followed by a separation token
(\texttt{<s>}), and finally the \textit{target sequence} (\circled{2}) that matches
the original input sequence. This is inspired by the concept of
\textit{content planning}, originating from natural language
generation~\cite{reiter1997building}. In content planning, a
high-level plan is first outlined, specifying the discourse structure
of the content to be generated, and then lexical realization is
performed to generate the text.

\subsubsection{Edit Plan}
\label{sec:edit-action}
The edit plan entails the specific edit operations that are needed to recover the original input sequence. For example, in
Figure~\ref{fig:pretrain-model}, the input sequence: ``\texttt{@param}
users List of user objects'' is corrupted by masking ``users'' and removing token ``user'':
``\texttt{@param} \texttt{[MASK]} List of objects''.
With this, a model must first reason about the fact that \texttt{[MASK]} in the
corrupted input sequence needs to be replaced with ``users'' and ``user'' should be inserted between ``of'' and ``objects'' when producing the target sequence.
To construct the sequence of edit operations, we closely follow the format proposed by \citet{PanthaplackelETAL20Learning}:
\begin{center}
\texttt{<Operation> [span of tokens] <OperationEnd>}
\end{center}
Here, \texttt{<Operation>} is either \Insert or \Delete.
We also include the \texttt{Replace} operation, with a slightly different structure (since both the old content to be replaced as well as the new content to replace it with must be specified):
\begin{center}
\texttt{<ReplaceOld> [span of old tokens] \\
<ReplaceNew> [span of new tokens] <ReplaceEnd>}
\end{center}
To determine the specific edit operations for a given example, we use difflib\footnote{\url{https://docs.python.org/3/library/difflib.html}} to compute the optimal set of edits needed to transform the corrupted input sequence into the original input sequence.
Multiple edit operations are placed in the same order as the span of tokens under editing appears in the input sequence (for example, the edit plan in Figure~\ref{fig:pretrain-model} consists of two edit operations).

\subsubsection{Target Sequence}
\label{sec:target-sequence}
One might ask whether we could simply apply the sequence of edit
operations in the generated edit plan to the corrupted input sequence
directly to recover the original input sequence
\textit{heuristically}. For example, if we align ``\texttt{<ReplaceOld>
	\texttt{[MASK]} <ReplaceNew> user <ReplaceOld>}'' with a corrupted input
sequence ``\texttt{@param} \texttt{[MASK]} List of user objects'', it is very
clear that all we need to do is replace \texttt{[MASK]} with ``user''and no
additional generation is needed. However,
there are two main issues with this. First, not all operations will be
specified in a deterministic manner.
For example, if the edit plan is ``\texttt{<Insert> user
<InsertEnd>}'', it is not clear where the new token ``user'' should be
added to.
Second, the generated edit plan does not correspond to contiguous
output tokens since it consists of fragmented information (edit
operations and token spans) rather than a complete sentence.  
As a result, neural language models may fail to generate correct edit
plans due to their lack of language properties such as fluency and
coherency~\cite{PanthaplackelETAL20Learning}.

Therefore, we need an additional step for \textit{learning} to apply
edits while simultaneously maintaining fluency and coherency.
For this reason, once
the edit plan is outlined as a sequence of edit operations, the target sequence (which is expected to recover the original input sequence) must also be generated: ``\texttt{@param} users List of user objects''. The decoder generates
tokens in a left-to-right manner, meaning that when generating a token
at a given timestep, it is aware of all tokens generated in previous
timesteps. So, when generating the target sequence, the decoder
can exploit the sequence of edits that was generated in the edit plan
earlier.
In this way, the model can reason the edits and the generation simultaneously.

%


\begin{table}[t]
    \begin{center}
        \caption{\UseMacro{TCap-span-stats}}
        \vspace{-10pt}
        \begin{tabular}{l | c  c }
            \toprule
             & \textbf{\UseMacro{THead-PL}}
             & \textbf{\UseMacro{THead-NL}}
            \\
            \midrule
            \UseMacro{THead-delete-prob}
             & \UseMacro{metrics-delete-prob-PL}
             & \UseMacro{metrics-delete-prob-NL}
            \\
            \UseMacro{THead-insert-prob}
             & \UseMacro{metrics-insert-prob-PL}
             & \UseMacro{metrics-insert-prob-NL}
            \\
            \UseMacro{THead-replace-prob}
             & \UseMacro{metrics-replace-prob-PL}
             & \UseMacro{metrics-replace-prob-NL}
            \\
            \UseMacro{THead-mean-span-length}
             & \UseMacro{metrics-mean-span-length-PL}
             & \UseMacro{metrics-mean-span-length-NL}
            \\
            \UseMacro{THead-mean-span-count}
             & \UseMacro{metrics-mean-span-count-PL}
             & \UseMacro{metrics-mean-span-count-NL}
            \\
            \bottomrule
        \end{tabular}
        \label{tab:span-stats}
    \end{center}
\end{table}

\subsubsection{Noising Functions}
To support learning across a diverse set of edit actions during \pretraining, we consider multiple noising functions for corrupting the input sequence:
1) randomly masking spans with the special \texttt{[MASK]}
token which requires the model to replace it with the correct spans,
2) inserting \texttt{[MASK]} token at random positions which requires the
model to identify the useless spans and delete them and 3) deleting
spans of tokens in the input sequence which requires the model
pinpoint the position and add back the missing pieces.

\vspace{-6pt}
\subsection{\Pretraining Data}
\label{sub:pretrain_data}
\subsubsection{Data Collection}

Following prior work, we \pretrain \MOEDIT on large amounts of source code and natural language comments from the CodeSearchNet~\cite{HusainETAL19Codesearchnet} dataset which consists of functions of six programming languages (Java, Python, Ruby, Php, Go and JavaScript) together with the natural language comments.
CodeSearchNet is widely used to \pretrain large language models, such as \CodeTf~\cite{wang2021codet5} and UniXcoder~\cite{GuoETAL22Unixcoder}.
We use the training set of the processed CodeSearchNet dataset
provided by~\citet{GuoETAL22Unixcoder} which contains
\UseMacro{pl-pretrain-number-initial}~million programming languages
code snippets (functions/methods) and
\UseMacro{nl-pretrain-number-initial}~million natural language
comments.


\begin{table}
    \begin{center}
        \caption{\UseMacro{TCap-pretrain-dataset}}
        \vspace{-10pt}
        \begin{tabular}{l | c  c }
            \toprule

             & \textbf{\UseMacro{THead-code}}
             & \textbf{\UseMacro{THead-nl}}
            \\
            \midrule
            \UseMacro{THead-examples-count}
             & \UseMacro{ds-pretrain-code-examples-count}
             & \UseMacro{ds-pretrain-nl-examples-count}
            \\
            \UseMacro{THead-buggy-avg-tokens-count}
             & \UseMacro{ds-pretrain-code-buggy-avg-tokens-count}
             & \UseMacro{ds-pretrain-nl-buggy-avg-tokens-count}
            \\
            \UseMacro{THead-edit-avg-tokens-count}
             & \UseMacro{ds-pretrain-code-edit-avg-tokens-count}
             & \UseMacro{ds-pretrain-nl-edit-avg-tokens-count}
            \\
            \bottomrule
        \end{tabular}
        \label{tab:pretrain-dataset}
    \end{center}
\end{table}

\subsubsection{Data Preparation}

To enable \MOEDIT to capture common edit patterns, we want the
\pretraining dataset to reflect the common activities conducted by
software developers.
Specifically, in the \pretraining dataset,
the probability of each edit operations applied to the spans in the input sequence and
the length (number of tokens) of the corrupted span should be consistent with the distributions and sizes of real-world edits in downstream editing tasks.

To this end, we collect statistics for source code edits from the training sets of the \bugfix and \codereview downstream tasks and statistics for natural language edits from the {\cupdate}'s training set.
As shown in Table~\ref{tab:span-stats}, we collect the probability of each edit operation
(insert, delete and replace) to be performed on a span; the average number of tokens in each span that is edited;
and the average number of spans that are edited in
each input sequence.
For each example in the \pretraining dataset, we then uniformly sample the spans and the edit operations that should be applied in accordance with the statistics collected from the downstream datasets.

Similar to \CodeTf~\cite{wang2021codet5}, we use the
RoBERTa~\cite{liu2019roberta} tokenizer to tokenize all sequences
(input, edit plan, target). More concretely, the tokenizer splits
words in the sequence into\XComment{ \textit{subwords}, constituting
the} \textit{tokens} (subwords) that are used by the model.  Moreover,
we remove input sequences that are shorter than 3 tokens and longer
than 512 tokens after tokenization which leave us with
\UseMacro{pl-pretrain-number}~million programming language code
snippets and \UseMacro{nl-pretrain-number}~million natural language
comments.  
This is because too short inputs are usually incomplete and \CodeTf is designed to only handle sequence of length 512.
Table~\ref{tab:pretrain-dataset} presents the statistics of
the \pretraining dataset.

\subsection{\Pretraining Setup}
\label{sub:pretrain_setup}

\paragraph{Model Architecture}
\MOEDIT consists of 12 encoder and decoder layers, 12 attention heads, and a hidden dimension size of 768. The total
number of parameters is 223M. Model parameters are initialized from the \CodeTf-base model, and we further \pretrain it on the CodeSearchNet \pretraining dataset (Section~\ref{sub:pretrain_data}) using our proposed objective (Section~\ref{sub:pretraining}).

\paragraph{Training}
We implement \MOEDIT using PyTorch 1.9.0 and use 16 NVidia 1080-TI GPUs, Intel(R) Xeon(R) CPU
E5-2620 v4 @ 2.10GHz for \pretraining for 4 days.
For fine-tuning, we run the experiments on 4
NVidia 1080-TI GPUs, Intel(R) Xeon(R) CPU E5-2620 v4 @ 2.10GHz with
the same hyper-parameters as \PLBART.


\begin{table}[t]
    \begin{center}
        \caption{\UseMacro{TCap-models-copy-pct-results}}
        \vspace{-10pt}
        \begin{tabular}{l | c  c  c  c }
            \toprule
            \textbf{\UseMacro{THead-models}}
             & \textbf{\UseMacro{THead-plbart}}
             & \textbf{\UseMacro{THead-CodeT5}}
             & \textbf{\UseMacro{THead-CoditT5}}
            \\
            \midrule
            \UseMacro{THead-bf-small1}
             & \UseMacro{results-bf-small-plbart-COPY-PCT}
             & \UseMacro{results-bf-small-CodeT5-COPY-PCT}
             & \UseMacro{results-bf-small-CoditT5-COPY-PCT}
            \\
            \UseMacro{THead-bf-medium}
             & \UseMacro{results-bf-medium-plbart-COPY-PCT}
             & \UseMacro{results-bf-medium-CodeT5-COPY-PCT}
             & \UseMacro{results-bf-medium-CoditT5-COPY-PCT}
            \\
            \UseMacro{THead-comment-update} (clean)
             & \UseMacro{results-comment-update-plbart-COPY-PCT}
             & \UseMacro{results-comment-update-CodeT5-COPY-PCT}
             & \UseMacro{results-comment-update-CoditT5-COPY-PCT}
            \\
            \UseMacro{THead-comment-update-full}
             & \UseMacro{results-comment-update-full-plbart-COPY-PCT}
             & \UseMacro{results-comment-update-full-CodeT5-COPY-PCT}
             & \UseMacro{results-comment-update-full-CoditT5-COPY-PCT}
            \\
            \UseMacro{THead-code-review}
             & \UseMacro{results-code-review-plbart-COPY-PCT}
             & \UseMacro{results-code-review-CodeT5-COPY-PCT}
             & \UseMacro{results-code-review-CoditT5-COPY-PCT}
            \\
            \bottomrule
        \end{tabular}
        \label{tab:copy-pct-results}
    \end{center}
\end{table}

\section{Experimental Design}


\begin{table}[t]
  \begin{center}
    \caption{Statistics for the datasets used for downstream tasks.}
    \vspace{-10pt}
    \begin{tabular}{l | c  c  c  c  c }
      \toprule
      \textbf{\UseMacro{THead-task}}
       &
       & \textbf{\UseMacro{THead-train-examples-count}}
       & \textbf{\UseMacro{THead-valid-examples-count}}
       & \textbf{\UseMacro{THead-test-examples-count}}
      \\
      \midrule
      \multirow{2}{*}{\UseMacro{THead-comment-update}}
       & clean
       & \UseMacro{ds-comment-update-train-examples-count}
       & \UseMacro{ds-comment-update-valid-examples-count}
       & \UseMacro{ds-comment-update-test-examples-count}
      \\
       & full
       & \UseMacro{ds-comment-update-full-train-examples-count}
       & \UseMacro{ds-comment-update-full-valid-examples-count}
       & \UseMacro{ds-comment-update-full-test-examples-count}
      \\
      \midrule
      \multirow{2}{*}{\UseMacro{THead-bf-small}}
       & \UseMacro{ds-bf-small-dataset}
       & \UseMacro{ds-bf-small-train-examples-count}
       & \UseMacro{ds-bf-small-valid-examples-count}
       & \UseMacro{ds-bf-small-test-examples-count}
      \\
       & \UseMacro{ds-bf-medium-dataset}
       & \UseMacro{ds-bf-medium-train-examples-count}
       & \UseMacro{ds-bf-medium-valid-examples-count}
       & \UseMacro{ds-bf-medium-test-examples-count}
      \\
      \midrule
      \UseMacro{THead-code-review}
       & \UseMacro{-}
       & \UseMacro{ds-code-review-train-examples-count}
       & \UseMacro{ds-code-review-valid-examples-count}
       & \UseMacro{ds-code-review-test-examples-count}
      \\
      \bottomrule
    \end{tabular}
    \label{tab:dataset}
  \end{center}
\end{table}

To assess \MOEDIT and our proposed \pretraining objective, we \finetune the model on three software-related downstream tasks.
Note that during \finetuning, the model is still trained to generate the edit-based output sequence.
However, at test time, we discard the edit plan and take the generated target sequence as the final model output.
Namely, we use the generated sequence after the separation token \texttt{<s>} as model's prediction.

\subsection{Downstream Tasks}
\label{sec:downstream-task}

\paragraph{\Cupdate}
The task of \cupdate entails automatically updating a natural language
comment to reflect changes in the corresponding body of
code~\cite{PanthaplackelETAL20Learning}. For instance, in Example~2 
in Figure~\ref{tab:cup-example}, the old \texttt{@return} comment needs to
be revised based on the changes in the method. Instead of
directly returning the yaw Euler angle measured in radians, the
unit of the return value is changed to degrees in the new version, with the method call \texttt{Math.toDegrees()}.

\paragraph{Bug Fixing}
Given a \textit{buggy} code snippet, the task of \bugfix entails
generating a \textit{fixed} code snippet, which no longer contains the
bug~\cite{tufano2019learning}.

\paragraph{\Codereview}
Given a code snippet under review and a brief natural language
sentence prescribing code edits, \codereview requires automatically
generating the revised code snippet, which captures the recommended
changes~\cite{Tufano21Towards}. For example, in
Figure~\ref{fig:intro-example:1}, \texttt{emptyList()} should be
changed to \texttt{Collections.}\-\texttt{emptyList()} because the
reviewer suggests \textit{not} using static import.

\subsection{Data for Downstream Tasks}
\label{sec:downstream-ds}
We use datasets that have been established and previously used for
each of the three tasks.  The statistics of the datasets is shown in
Table~\ref{tab:dataset}.  Unlike \pretraining where the goal is to
recover the corrupted input sequences, during \finetuning, \MOEDIT is
trained to generate an edit plan for completing the downstream editing
task, that can be applied to a part of the input (e.g., old comment),
followed by the target sequence (e.g., new comment).

\paragraph{\Cupdate}
For this task, \citet{PanthaplackelETAL20Deep} has released a corpus
of Java method changes paired with changes in the corresponding
comments (spanning \texttt{@return}, \texttt{@param}, and summary
comments). This dataset also comes with a \textit{clean} subset of the
test set which was manually curated.
The input sequence used for \finetuning is formed by concatenating the old
comment and code edits. The code edits follow
the representation described in Section~\ref{sec:edit-action}, except that an additional \texttt{Keep} operation is included to denote spans that are left unchanged.

\paragraph{Bug Fixing}
We consider the Java BugFixPairs-Small (\bfs) and BugFixPairs-Medium
(\bfm) datasets, originally released by~\citet{tufano2019learning}.
\citet{ChakrabortyAndRay21Multi} supplemented these datasets with
additional context, namely natural language guidance from the
developer, and the method  where the patch should be applied. \bfs
contains shorter methods with a maximum token length 50, and \bfm
contains longer methods with up to 100 tokens in length.  The input
sequence used for \finetuning is formed with the buggy code, natural
language guidance, and code context.

\paragraph{Automated Code Review}
We use the \codereview dataset released by~\citet{Tufano21Towards},
which consists of Java methods (before and after the review) paired
with pull request comments, derived from pull request reviews on
GitHub and Gerrit.  To reduce the vocabulary size, they further
abstracted Java methods by replacing identifiers and literals with
special tokens.  In this work, we use the data with concrete tokens.
The input sequence used for \finetuning is formed using the code
snippet before review and the pull request comment from reviewers.

\subsection{Baselines}
\subsubsection{Generation Baselines}
\label{sub:generation_baselines}
We consider two large standard generation language models trained with
denoising autoencoding \pretraining objectives which are not
edit-based: \textbf{\PLBARTft} and \textbf{\CodeTf}. Both of these are
\finetuned to directly generate the target output sequence.
Furthermore, to better assess the value of actually \pretraining using
the proposed objective instead of simply \finetuning a model to
generate an edit-based output sequence, we also consider \finetuning
\CodeTf to generate the specialized edit-based output sequence
representation. We refer to this as \textbf{\MOEDITft}. We \finetune
each of these models using the same input context as \MOEDIT.

\subsubsection{Task-Specific Baselines}
We additionally compare against the state-of-the-art models for each of the downstream tasks. 

For \cupdate, the state-of-the-art model is \textbf{\editmodel}, which
entails Recurrent Neural Network (RNN) based encoders for representing
the old comment and code edits, and an RNN-based decoder for decoding
edits.  These edits are parsed at test time and reranked based on
similarity to the old comment and likelihood based on a comment
generation model.

For \bugfix, the state-of-the-art model is essentially \PLBARTft
\finetuned on the \bfs and \bfm to generate the fixed
code~\cite{ChakrabortyAndRay21Multi}.

For \codereview, no baselines are available for the specific version
of the dataset we used with concrete identifiers and literals (rather
than the one with abstracted identifiers and literals).  Therefore, we
rely on those described in Section~\ref{sub:generation_baselines} and
establish new baselines for this version of the dataset.

\subsection{Evaluation Metrics}

For \cupdate, we report performance on the same metrics that have been
used previously to benchmark models for this
task~\cite{PanthaplackelETAL20Learning}. This includes: xMatch
(whether the model prediction \textit{exactly matches} the ground
truth), common metrics that measure lexical overlap for evaluating
text generation (BLEU-4~\footnote{We measure 1$\sim$4-gram overlap and
compute the average.}~\cite{papineni2002bleu} and
METEOR~\cite{banerjee2005meteor}), and common metrics for measuring
text editing (GLEU~\cite{napoles2015ground} and
SARI~\cite{xu2016optimizing}).  For \bugfix, we use xMatch, as done in
prior work~\cite{ChakrabortyAndRay21Multi}. For \codereview, we report
performance on xMatch and BLEU-4, which have been used previously to
benchmark models for this task~\cite{Tufano21Towards}.

\section{Evaluation}

We organize our evaluation around three main research questions:

\RQ{1}{How does our edit-based model, \MOEDIT, compare to generation
	and task-specific baselines for edit-related tasks?}

\RQ{2}{Does our proposed \pretraining objective help a model in better reasoning about and performing edits?}

\RQ{3}{Can a standard generation model complement \MOEDIT by integrating the two models?}

\subsection{Comparing \MOEDIT to Baselines}

\begin{figure}[t]
  \centering
  \begin{tabular}{l}
    \toprule
    \underline{\small Before Editing:}     \\
    {
\begin{lstlisting}[language=java-pretty, numbers=none]
public HashConfigurationBuilder capacityFactor (float capacityFactor) {
    if ( numSegments < 0 )
	throw new IllegalArgumentException ("capacityFactor must be positive");
    this.capacityFactor = capacityFactor ;
    return this;
}
\end{lstlisting}}
    \\
    \underline{\small Reviewer's Comment:} \\
    {\small typo: capacityFactor instead of numSegments}
    \\
    \underline{\small CodeT5:}             \\
    {
\begin{lstlisting}[language=java-pretty, numbers=none]
public HashConfigurationBuilder capacityFactor(float capacityFactor) {
    this.capacityFactor = capacityFactor;
    return this;
}
\end{lstlisting}}
    \\
    \underline{\small \MOEDIT:}            \\
    {
\begin{lstlisting}[language=java-pretty, numbers=none]
public HashConfigurationBuilder capacityFactor (float capacityFactor) {
    if ( capacityFactor < 0 )
	throw new IllegalArgumentException ("capacityFactor must be positive") ;
    this.capacityFactor = capacityFactor;
    return this;
}
\end{lstlisting}}
    \\
    \bottomrule
  \end{tabular}
  \caption{Comparing the output of \CodeTf and \MOEDIT for a \codereview example. \CodeTf generates incorrect output that drastically deviates from the input code while \MOEDIT generates the correct output, performing only relevant edits.}
  \label{tab:qual-example}
\end{figure}


\begin{table*}[h]
    \begin{center}
        \caption{\UseMacro{TCap-comment-update-models-results}}
        \vspace{-10pt}
        \begin{tabular}{l | c  c  c  c  c  c }
            \toprule
            \textbf{\UseMacro{THead-models}}
             & \textbf{\UseMacro{THead-Acc-top-1}}
             & \textbf{\UseMacro{THead-BLEU}}
             & \textbf{\UseMacro{THead-METEOR}}
             & \textbf{\UseMacro{THead-GLEU}}
             & \textbf{\UseMacro{THead-SARI}}
            \\
            \midrule
            \UseMacro{THead-EditModel}
             & \UseMacro{results-comment-update-EditModel-Acc-top-1}
             & \UseMacro{results-comment-update-EditModel-BLEU}
             & \UseMacro{results-comment-update-EditModel-METEOR}
             & \UseMacro{results-comment-update-EditModel-GLEU}
             & \UseMacro{results-comment-update-EditModel-SARI}
            \\
            \UseMacro{THead-plbart}
             & \UseMacro{results-comment-update-plbart-Acc-top-1}
             & \UseMacro{results-comment-update-plbart-BLEU}$^{\gamma}$
             & \UseMacro{results-comment-update-plbart-METEOR}
             & \UseMacro{results-comment-update-plbart-GLEU}
             & \UseMacro{results-comment-update-plbart-SARI}
            \\
            \UseMacro{THead-CodeT5}
             & \UseMacro{results-comment-update-CodeT5-Acc-top-1}
             & \UseMacro{results-comment-update-CodeT5-BLEU}$^{\alpha}$
             & \UseMacro{results-comment-update-CodeT5-METEOR}
             & \UseMacro{results-comment-update-CodeT5-GLEU}$^{\beta}$
             & \UseMacro{results-comment-update-CodeT5-SARI}
            \\
            \UseMacro{THead-CoditT5-ft}
             & \UseMacro{results-comment-update-CoditT5-ft-Acc-top-1}
             & \UseMacro{results-comment-update-CoditT5-ft-BLEU}$^{\gamma}$
             & \UseMacro{results-comment-update-CoditT5-ft-METEOR}
             & \UseMacro{results-comment-update-CoditT5-ft-GLEU}$^{\beta}$
             & \UseMacro{results-comment-update-CoditT5-ft-SARI}$^{\epsilon}$$^{\eta}$
            \\
                    \UseMacro{THead-CoditT5}
             & \UseMacro{results-comment-update-CoditT5-Acc-top-1}$^{\chi}$
             & \UseMacro{results-comment-update-CoditT5-BLEU}
             & \UseMacro{results-comment-update-CoditT5-METEOR}
             & \UseMacro{results-comment-update-CoditT5-GLEU}
             & \UseMacro{results-comment-update-CoditT5-SARI}$^{\delta}$$^{\epsilon}$
            \\
                \midrule
                \UseMacro{THead-CoditT5-CodeT5-rerank}
             & \textbf{\UseMacro{results-comment-update-CoditT5-CodeT5-rerank-Acc-top-1}}
             & \textbf{\UseMacro{results-comment-update-CoditT5-CodeT5-rerank-BLEU}}
             & \textbf{\UseMacro{results-comment-update-CoditT5-CodeT5-rerank-METEOR}}
             & \textbf{\UseMacro{results-comment-update-CoditT5-CodeT5-rerank-GLEU}}
             & \UseMacro{results-comment-update-CoditT5-CodeT5-rerank-SARI}$^{\delta}$$^{\eta}$
            \\
                \UseMacro{THead-CodeT5-CoditT5-rerank}
             & \UseMacro{results-comment-update-CodeT5-CoditT5-rerank-Acc-top-1}$^{\chi}$
             & \UseMacro{results-comment-update-CodeT5-CoditT5-rerank-BLEU}$^{\alpha}$
             & \UseMacro{results-comment-update-CodeT5-CoditT5-rerank-METEOR}
             & \UseMacro{results-comment-update-CodeT5-CoditT5-rerank-GLEU}
             & \textbf{\UseMacro{results-comment-update-CodeT5-CoditT5-rerank-SARI}}
            \\
            \bottomrule
        \end{tabular}
        \label{tab:comment-update-results}
    \end{center}
\end{table*}


\begin{table*}[h]
    \begin{center}
        \caption{\UseMacro{TCap-comment-update-full-models-results}}
        \vspace{-10pt}
        \begin{tabular}{l | c  c  c  c  c  c }
            \toprule
            \textbf{\UseMacro{THead-models}}
             & \textbf{\UseMacro{THead-Acc-top-1}}
             & \textbf{\UseMacro{THead-BLEU}}
             & \textbf{\UseMacro{THead-METEOR}}
             & \textbf{\UseMacro{THead-GLEU}}
             & \textbf{\UseMacro{THead-SARI}}
            \\
            \midrule
            \UseMacro{THead-EditModel}
             & \UseMacro{results-comment-update-full-EditModel-Acc-top-1}
             & \UseMacro{results-comment-update-full-EditModel-BLEU}
             & \UseMacro{results-comment-update-full-EditModel-METEOR}
             & \UseMacro{results-comment-update-full-EditModel-GLEU}
             & \UseMacro{results-comment-update-full-EditModel-SARI}
            \\
            \UseMacro{THead-plbart}
             & \UseMacro{results-comment-update-full-plbart-Acc-top-1}
             & \UseMacro{results-comment-update-full-plbart-BLEU}$^{\hbar}$$^{\iota}$
             & \UseMacro{results-comment-update-full-plbart-METEOR}
             & \UseMacro{results-comment-update-full-plbart-GLEU}
             & \UseMacro{results-comment-update-full-plbart-SARI}
            \\
                        \UseMacro{THead-CodeT5}
             & \UseMacro{results-comment-update-full-CodeT5-Acc-top-1}
             & \UseMacro{results-comment-update-full-CodeT5-BLEU}$^{\alpha}$
             & \UseMacro{results-comment-update-full-CodeT5-METEOR}
             & \UseMacro{results-comment-update-full-CodeT5-GLEU}
             & \UseMacro{results-comment-update-full-CodeT5-SARI}
            \\
                        \UseMacro{THead-CoditT5-ft}
             & \UseMacro{results-comment-update-full-CoditT5-ft-Acc-top-1}$^{\delta}$$^{\gamma}$
             & \UseMacro{results-comment-update-full-CoditT5-ft-BLEU}
             & \UseMacro{results-comment-update-full-CoditT5-ft-METEOR}
             & \UseMacro{results-comment-update-full-CoditT5-ft-GLEU}$^{\epsilon}$
             & \UseMacro{results-comment-update-full-CoditT5-ft-SARI}$^{\eta}$
            \\
                    \UseMacro{THead-CoditT5}
             & \UseMacro{results-comment-update-full-CoditT5-Acc-top-1}$^{\delta}$
             & \UseMacro{results-comment-update-full-CoditT5-BLEU}$^{\beta}$$^{\iota}$
             & \UseMacro{results-comment-update-full-CoditT5-METEOR}$^{\chi}$
             & \UseMacro{results-comment-update-full-CoditT5-GLEU}$^{\epsilon}$
             & \UseMacro{results-comment-update-full-CoditT5-SARI}
            \\
                        \midrule
                        \UseMacro{THead-CoditT5-CodeT5-rerank}
             & \textbf{\UseMacro{results-comment-update-full-CoditT5-CodeT5-rerank-Acc-top-1}}$^{\gamma}$
             & \textbf{\UseMacro{results-comment-update-full-CoditT5-CodeT5-rerank-BLEU}}$^{\alpha}$
             & \textbf{\UseMacro{results-comment-update-full-CoditT5-CodeT5-rerank-METEOR}}
             & \textbf{\UseMacro{results-comment-update-full-CoditT5-CodeT5-rerank-GLEU}}
             & \UseMacro{results-comment-update-full-CoditT5-CodeT5-rerank-SARI}
            \\
                        \UseMacro{THead-CodeT5-CoditT5-rerank}
             & \UseMacro{results-comment-update-full-CodeT5-CoditT5-rerank-Acc-top-1}
             & \UseMacro{results-comment-update-full-CodeT5-CoditT5-rerank-BLEU}$^{\beta}$$^{\hbar}$
             & \UseMacro{results-comment-update-full-CodeT5-CoditT5-rerank-METEOR}$^{\chi}$
             & \UseMacro{results-comment-update-full-CodeT5-CoditT5-rerank-GLEU}
             & \textbf{\UseMacro{results-comment-update-full-CodeT5-CoditT5-rerank-SARI}}$^{\eta}$
            \\
            \bottomrule
        \end{tabular}
        \label{tab:comment-update-full-results}
    \end{center}
\end{table*}


\begin{table}[t]
  \begin{center}
    \caption{\UseMacro{TCap-bf-models-results}}
    \vspace{-10pt}
    \begin{tabular}{l | c  c }
      \toprule
      \multirow{2}{*}{
        \textbf{\UseMacro{THead-models}}}
       & \multicolumn{2}{c}{\textbf{\UseMacro{THead-Acc-top-1}}}
      \\
       & \textbf{\bfs}                                                                   & \textbf{\bfm} \\
      \midrule
      \UseMacro{THead-plbart}
       & \UseMacro{results-bf-small-plbart-Acc-top-1}
       & \UseMacro{results-bf-medium-plbart-Acc-top-1}

      \\
      \UseMacro{THead-CodeT5}
       & \UseMacro{results-bf-small-CodeT5-Acc-top-1}
       & \UseMacro{results-bf-medium-CodeT5-Acc-top-1}
      \\
      \UseMacro{THead-CoditT5-ft}
       & \UseMacro{results-bf-small-CoditT5-ft-Acc-top-1}
       & \UseMacro{results-bf-medium-CoditT5-ft-Acc-top-1}$^{\alpha}$
      \\
      \UseMacro{THead-CoditT5}
       & \UseMacro{results-bf-small-CoditT5-Acc-top-1}
       & \UseMacro{results-bf-medium-CoditT5-Acc-top-1}$^{\alpha}$
      \\
      \midrule
      \UseMacro{THead-CoditT5-CodeT5-rerank}
       & \textbf{\UseMacro{results-bf-small-CoditT5-CodeT5-rerank-Acc-top-1}}
       & \textbf{\UseMacro{results-bf-medium-CoditT5-CodeT5-rerank-Acc-top-1}}$^{\beta}$
      \\
      \UseMacro{THead-CodeT5-CoditT5-rerank}
       & \UseMacro{results-bf-small-CodeT5-CoditT5-rerank-Acc-top-1}
       & \UseMacro{results-bf-medium-CodeT5-CoditT5-rerank-Acc-top-1}$^{\beta}$
      \\
      \bottomrule
    \end{tabular}
    \label{tab:bf-results}
  \end{center}
\end{table}


\begin{table}[t]
    \begin{center}
        \caption{\UseMacro{TCap-code-review-models-results}}
        \vspace{-10pt}
        \begin{tabular}{l | c  c  c }
            \toprule
            \textbf{\UseMacro{THead-models}}
             & \textbf{\UseMacro{THead-Acc-top-1}}
             & \textbf{\UseMacro{THead-BLEU}}
            \\
            \midrule
            \UseMacro{THead-plbart}
             & \UseMacro{results-code-review-plbart-Acc-top-1}
             & \UseMacro{results-code-review-plbart-BLEU}
            \\
            \UseMacro{THead-CodeT5}
             & \UseMacro{results-code-review-CodeT5-Acc-top-1}
             & \UseMacro{results-code-review-CodeT5-BLEU}
            \\
            \UseMacro{THead-CoditT5-ft}
             & \UseMacro{results-code-review-CoditT5-ft-Acc-top-1}$^{\alpha}$
             & \UseMacro{results-code-review-CoditT5-ft-BLEU}$^{\beta}$
            \\
            \UseMacro{THead-CoditT5}
             & \UseMacro{results-code-review-CoditT5-Acc-top-1}$^{\alpha}$
             & \UseMacro{results-code-review-CoditT5-BLEU}$^{\beta}$
            \\
            \midrule
            \UseMacro{THead-CoditT5-CodeT5-rerank}
             & \UseMacro{results-code-review-CoditT5-CodeT5-rerank-Acc-top-1}
             & \textbf{\UseMacro{results-code-review-CoditT5-CodeT5-rerank-BLEU}}$^{\chi}$
            \\
            \UseMacro{THead-CodeT5-CoditT5-rerank}
             & \textbf{\UseMacro{results-code-review-CodeT5-CoditT5-rerank-Acc-top-1}}
             & \UseMacro{results-code-review-CodeT5-CoditT5-rerank-BLEU}$^{\chi}$
            \\
            \bottomrule
        \end{tabular}
        \label{tab:code-review-results}
    \end{center}
\end{table}

\begin{figure}[t]
    \centering
    \renewcommand{\arraystretch}{1} 
    \begin{tabular}{l}
        \toprule

        Example 1                                                                                                                                 \\
        \underline{\small Before Editing:}                                                                                                        \\
        {
\begin{lstlisting}[language=java-pretty, numbers=none]
protected boolean isProcessed(ChronicleLogOffsetTracker tracker, long offset) {
    long last = tracker.readLastCommittedOffset();
    return ( last > 0 ) && ( last >= offset );
}
\end{lstlisting}}
        \\
        \underline{\small Reviewer's Comment:}                                                                                                    \\
        {\small No need for parentheses.}
        \\
        \underline{\small Edit plan}                                                                                                              \\
        {\scriptsize \texttt{$\langle$Delete$\rangle$ ( $\langle$Delete\_End$\rangle$ $\langle$Delete$\rangle$ ) $\langle$Delete\_End$\rangle $}} \\
        \underline{\small Target sequence:}                                                                                                       \\
        {
\begin{lstlisting}[language=java-pretty, numbers=none]
protected boolean isProcessed(ChronicleLogOffsetTracker tracker, long offset) {
    long last = tracker.readLastCommittedOffset();
    return last > 0 && last >= offset;
}
\end{lstlisting}}
        \\
        \midrule
        Example 2                                                                                                                                 \\
        \underline{\small Before Editing:}                                                                                                        \\
        {
\begin{lstlisting}[language=java-pretty, numbers=none]
public Builder setDataSize(Estimate dataSize) { 
    this.dataSize = requireNonNull(dataSize, "dataSize can not be null");
    return this; 
}
\end{lstlisting}}
        \\
        \underline{\small Reviewer's Comment}                                                                                                     \\
        {\small you don't validate in other builders method (and you don't have to)}                                                              \\
        \underline{\small Edit plan}                                                                                                              \\
        {\scriptsize \texttt{$\langle$Delete$\rangle$ equireNonNull(dataSize, "dataSize can not be null"); $\langle$Delete\_End$\rangle$}}        \\
        \underline{\small Target sequence:}                                                                                                       \\
        {
\begin{lstlisting}[language=java-pretty, numbers=none]
public Builder setDataSize(Estimate dataSize) {
    this.dataSize = dataSize; 
    return this; 
}
\end{lstlisting}}
        \\

        \bottomrule
    \end{tabular}
    \caption{Examples for \codereview for which \MOEDIT generated ambiguous or erroneous edit plans but still managed to generate the correct target sequences.}\label{fig:new-objective-code-review-example}
\end{figure}

We present results in Tables~\ref{tab:comment-update-results}-\ref{tab:code-review-results}.
Note that the results shown in the last two rows in each of the tables are explained later in Section~\ref{sub:rerank}. We perform statistical significance testing using bootstrap
tests~\cite{Berg-KirkpatrickETAL12Empirical} with confidence level 95\%.
\\

\noindent\fbox{%
	\parbox{.96\columnwidth}{%
		\textbf{RQ1:} How does our edit-based model, \MOEDIT, compare to generation and task-specific baselines for edit-related tasks?
	}%
}\\

We find that \MOEDIT (and most of the \pretrained models) drastically outperforms
\citet{PanthaplackelETAL20Learning} (a non-\pretrained model) across metrics for \cupdate. This demonstrates the value of large language model \pretrained on vast amounts of data using unsupervised \pretraining objectives.

Next, across all three tasks, \MOEDIT achieves higher performance than the two standard generation-based \pretrained models, significantly outperforming \PLBARTft and \CodeTf for most of the metrics, highlighting the benefit of explicitly modeling edits for these editing tasks.
In fact, \MOEDITft, which explicitly models edits only during \finetuning rather than \pretraining, outperforms \CodeTf on edit-based metrics (xMatch, SARI).
This further underlines the utility of the edit-based output sequence representation that we developed.

Nonetheless, across most metrics, \MOEDIT still outperforms \MOEDITft, which is not \pretrained using the \pretraining objective but uses the same edit-based output sequence representation during \finetuning. This demonstrates the importance of actually \pretraining with this representation rather than relying on \finetuning alone.

\subsection{Evaluating our \Pretraining Objective}


\begin{table}[t]
    \begin{center}
        \caption{Percentages of target sequence generated by \MOEDIT being consistent with the edit plan.}
        \vspace{-10pt}
        \begin{tabular}{l | c  }
            \toprule
            \textbf{Datasets}
             & \textbf{Is Consistent (\%)}
            \\
            \midrule
            \UseMacro{THead-bf-small1}
             & 92\%
            \\
            \UseMacro{THead-bf-medium}
             & 88\%
            \\
            \UseMacro{THead-comment-update} (clean)
             & 87\%
            \\
            \UseMacro{THead-comment-update-full}
             & 85\%
            \\
            \UseMacro{THead-code-review}
             & 74\%
            \\
            \bottomrule
        \end{tabular}
        \label{tab:plan-edit-consistent}
    \end{center}
\end{table}

While we observe that \MOEDIT tends to achieve slightly lower performance than \CodeTf on generation-based metrics (BLEU-4, METEOR) for two of the tasks, we find that it significantly outperforms other metrics which capture whether the correct edits are generated, such as xMatch and GLEU and SARI for \cupdate.
This suggests that \MOEDIT is indeed better at \textit{editing}.
By inspecting the outputs of the two models, we find that \CodeTf tends to make drastic and unnecessary edits while \MOEDIT appears to be better at making more fine-grained edits.
For example, in Figure~\ref{tab:qual-example}, \CodeTf generates output that completely discards critical statements in the code, whereas \MOEDIT is able to correctly localize the part of the input code that needs to be changed and make editions properly.
We attribute this to the fact that \CodeTf is not designed to reason about edits while \MOEDIT is.
We further evaluate the influence of our proposed \pretraining objective on this editing capability.\\

\noindent\fbox{%
	\parbox{.96\columnwidth}{%
		\textbf{RQ2:} Does our proposed \pretraining objective help a model in better reasoning about and performing edits?
	}%
}\\

First, we compare how often \MOEDIT naively copies the input content without actually performing any edits, to two \pretrained models which use generation-based \pretraining objectives.
We report the percentages in Table~\ref{tab:copy-pct-results}.
By copying substantially less often than the PLBART and \CodeTf, we find that \MOEDIT learns to more frequently perform edits with our proposed edit-based \pretraining objective which indicates it is suitable for editing tasks.

\MOEDIT's decoder is encouraged to generate a target sequence that follows the outlined edit plan; however, we do not constrain the decoder in any way to do this.\footnote{We do not want potential errors in the edit plan to propagate to the target sequence.}
Nonetheless, we find that in the majority of cases (\UseMacro{coditT5-consistent-rate-low}-\UseMacro{coditT5-consistent-rate-high}),  the target sequence is consistent with the edit plan, as shown in Table~\ref{tab:plan-edit-consistent}.
More concretely, the target sequence generally resembles what would be produced if the edit operations in the edit plan were applied to the original content.
This suggests that the \pretraining objective does in fact guide the model in reasoning about edits.

For cases in which there is ambiguity or errors in the edit plan, we find that \MOEDIT still often manages to generate the correct target sequence, by disregarding unreasonable edits or disambiguating ambiguous edits. We show two examples in \codereview in
Figure~\ref{fig:new-objective-code-review-example} with the Java method
before review, the generated edit plan, and the generated target sequence.
In Example 1, the edit plan is ambiguous since there are multiple instances of ``('' and it does not specify which one(s) should be deleted. However, the generated target sequence is correct, as the model was able to correctly reason about the most appropriate edit locations.
In Example 2, the edit plan is imprecise and blindly following this plan would result in syntactically incorrect code, but the model still managed to perform the correct edits and produced valid output by ignoring the fallacious edit.
Overall, we find that both components of the edit-based output sequence representation used in the \pretraining objective (edit plan and target sequence) are critical.
\subsection{Integrating \MOEDIT and \PLBART}
\label{sub:rerank}

\MOEDIT is designed to complement a generation model by providing more
explicit guidance for edits. However, a model that is trained to
generate edits can struggle with coherence and fluency since it is not
actually trained to generate consecutive
text~\cite{PanthaplackelETAL20Learning}. By\XComment{ also} including
the generation of the target sequence in the \pretraining objective,
we do mitigate this to some extent, even when there are ambiguities or
errors in the edit plan.  However, there appears to be a trade-off
between\XComment{ being able to perform} performing the correct edits
while maintaining performance with respect to generation metrics.
More specifically, in
Tables~\ref{tab:comment-update-results}-\ref{tab:code-review-results},
\MOEDIT outperforms \CodeTf with respect to xMatch (and SARI for
\cupdate), but underperforms with respect to BLEU-4. To exploit the
slight superiority of \CodeTf in this respect, we consider
incorporating \CodeTf into our approach.\\

\noindent\fbox{%
	\parbox{.96\columnwidth}{%
		\textbf{RQ3}: Can a pure generation model complement \MOEDIT by integrating the two models?
	}%
}\\

\begin{figure}[t]
  \centering
  \begin{tabular}{l}
    \toprule

    Example 1                                                                                          \\
    \underline{\small Buggy Code}                                                                      \\
    {
\begin{lstlisting}[language=java-pretty, numbers=none]
public List<TagVFilter> getFilters() { 
    if ((filters) == null ) { 
        filters = new ArrayList<TagVFilter>();
    }
    return filters; 
}
\end{lstlisting}}
    \\
    \underline{\small \MOEDIT:}                                                                        \\
    {
\begin{lstlisting}[language=java-pretty, numbers=none]
public List<TagVFilter> getFilters() { 
    if ((filters) == null ) { 
        filters = new ArrayList<TagVFilter>();
    }
    return new ArrayList(filters);
}
\end{lstlisting}}
    \\
    \underline{\small \MOEDITRerank:}                                                                  \\
    {
\begin{lstlisting}[language=java-pretty, numbers=none]
public List<TagVFilter> getFilters() { 
    if ((filters) == null ) { 
        filters = new ArrayList<TagVFilter>();
    }
    return new ArrayList<TagVFilter>(filters); 
}
\end{lstlisting}} \\

    \midrule

    Example 2                                                                                          \\
    {
\begin{lstlisting}[language=java-pretty, numbers=none]
/** @return double The yaw Euler angle. */
public double getRotY() {
    return mOrientation.getRotationY();
}

/** @return ? */
public double getRotY() {
    return Math.toDegrees(mOrientation.getRotationY());
}
\end{lstlisting}}
    \\
    {\small \underline{\PLBART}: @return double The yaw Euler angle.}
    \\
    {\small \underline{Reranked \PLBART}: @return double The yaw Euler angle in degrees.}              \\

    \bottomrule
  \end{tabular}
  \caption{Examples from \cupdate and \bugfix which demonstrate the impact of reranking.}
  \label{tab:cup-example}
\end{figure}

\subsubsection{Experimental Setup}

We combine the two models using simple likelihood-based reranking
strategies at test time (with no additional training). Namely, at test time, \MOEDIT and \CodeTf each generate 20 candidates using beam search.
While we have been only looking at the top one prediction for all previous experiments, we will consider all 20 candidates for reranking.
We compute a reranking score for each of these to essentially re-score them. The candidate which has the highest reranking score will be the final model prediction. We investigate two different reranking strategies:

\paragraph{\MOEDITRerank:}
To exploit the language-specific norms learned by \CodeTf, we rerank the candidates generated by \MOEDIT based on the probability score \CodeTf's language model assigns to the corresponding target sequences (namely after \texttt{<s>}).

We compute the length-normalized conditional log
probability score of \PLBART generating the target sequence, conditioned on the same
input:

\[ score = log(P(T|I)^{\frac{1}{N}}) \]
where $T$ is the target sequence, $I$ is the model's input, $N$ is
the length of $T$. We also length-normalize the log probability of the candidate, as scored by \MOEDIT, and then add the two probability scores together to obtain the reranking score.

\paragraph{\PLBARTRerank:} Conversely, we also rerank the output of \PLBART based on the likelihood of \MOEDIT, such that the generated sequence can be assessed in terms of explicit edits.
We first parse the output of \PLBART into the edit-based output sequence representation (as described in Section~\ref{sec:edit-action}) and then concatenate it with the model's output using \texttt{<s>}.
Then we compute the likelihood of \MOEDIT generating this sequence, conditioned on the same input. We then add the length-normalized log probability score of \MOEDIT with the score originally assigned by \PLBART (after length-normalizing and applying log).

\subsubsection{Results}
We provide results in the bottom two rows of Tables~\ref{tab:comment-update-results}-\ref{tab:code-review-results}. By reranking the output of \MOEDIT using \CodeTf, we are able to achieve improved performance on all the metrics including BLEU-4 across tasks (and the other generation-based metric, METEOR, for \cupdate). To illustrate this, consider Example~1 in Figure~\ref{tab:cup-example},
with a buggy code snippet and outputs corresponding to \MOEDIT before
and after reranking.
We observe that \MOEDIT correctly localizes the
bug and correctly identifies that the edit entails initializing an \CodeIn{ArrayList} in the return statement.
However, the generated target sequence is a defective code snippet which does not properly initialize an \CodeIn{ArrayList} with the correct type \CodeIn{TagVFilter}.
By leveraging \CodeTf's likelihood score, we are able to effectively filter out the defective prediction and obtain the correct output.

By reranking the output of \CodeTf using \MOEDIT, we see significant improvements with respect to \CodeTf on metrics that more directly evaluate whether the correct edits were performed, including xMatch as well as GLEU and SARI for \cupdate. This suggests that the edit-based and generation-based models are indeed complementary to one another. As a case
study, consider Example~2 in Figure~\ref{tab:cup-example}. \CodeTf
produces a sequence which simply copies the old comment,
without capturing the code changes. While this may be a likely comment
sequence, according to \PLBART's language model, copying without
applying any edits is not a likely edit plan to be generated for \MOEDIT.

By combining \MOEDIT and \CodeTf through reranking, we can further boost performance substantially across most metrics for all three tasks, outperforming the two models individually, and achieving new state-of-the-art.
\section{Limitations}

\MyPara{Other Programming Languages} The downstream editing tasks we
studied in this work are using Java.  Since \MOEDIT's \pretraining is
on the dataset consisting of six programming languages, we expect it
to also perform well on editing tasks in other programming languages,
but we leave empirically verifying this as future work.

\MyPara{Data Contamination} \MOEDIT is \pretrained on data collected
from open-source projects.  It is possible that similar examples in
\pretraining data exist in downstream tasks' test set.  While prior
work~\cite{brown2020language} has shown that data contamination may
have little impact on the performance of \pretrained models in natural
language processing tasks, future work can investigate this problem
for \pretrained models for software engineering.

\section{Related Work}
\label{sec:related}

In this section, we consider the most closely related work on learning
edits, large \pretrained models for code, \pretrained models for code edits and combining complementary
models.

\MyPara{Learning Edits} Prior work has studied learning edits in both
natural language and programming language.  We followed the approach
of explicitly representing edits as sequences with edit actions.  Our
edit representation is inspired by
\citet{PanthaplackelETAL20Learning,PanthaplackelETAL20Deep}, who
studied learning comment edits based on code edits.
\citet{BrodyETAL20CodeChanges,TarlowETAL20Learning,ChenETAL21PLUR,
  YaoETAL21Learning} represented code as ASTs (abstract syntax trees) and the code edits as edit actions over the AST nodes rather than tokens. We do not focus on editing
structured data (AST) as it can not be generalized to natural
language, and it can not be easily combined with large \pretrained
models which are primarily based on sequence of tokens.

Alternatively, edits can be encoded into vector representations (or
embeddings).  \citet{GuuETAL18EditingPrototypes} studied learning edit
embeddings for natural language generation in a prototype-then-edit
style.  \citet{YinETAL18RepresentEdits} studied learning code edits as
embeddings and then applying them to natural language insertion and
code bug fixing.  \citet{HashimotoETAL18RetrieveAndEdit} developed a
retrieve-and-edit framework for text-to-code generation, where the
edits are learned as parameters of a seq2seq model.
Similarly, \citet{LiETAL21Editsum} proposed a retrieve-and-edit framework for code summarization task where the model first learns an edit vector and then generate the revised summary conditioned on it.
Although learning edits as embeddings can be
effective for individual tasks, it is not suitable to be used in the
\pretraining \finetuning paradigm, because there is a large domain gap
between the edit embeddings learned on different tasks.   More over,
edit embeddings are less explainable compared to the explicit edit
representations we use.

Another line of work that carries out the idea of learning edits is
copying mechanism, including copying individual
tokens~\cite{VinyalsETAL15PointerNetworks,GuETAL16Copying} and
spans~\cite{ZhouETAL18Copying,PanthaplackelETAL21CopyThat}, which
helps the model to ``keep'' unchanged tokens and focus on generating
the edited part.
\citet{Logan21Fruit} built a T5-based model to update the existing articles based on the given new evidence. The model is trained to output a \textit{copy} token instead of the copied sentence and a special \textit{reference} token before the updated text  which identifies the evidence to support the update.
\citet{DingETAL20Patching} trained the model to emit
pointers that indicate the positions for editions and new tokens
to be inserted at the same time.
Similarly, \citet{TarlowETAL20Learning, ChenETAL21PLUR} augmented the transformer-based decoder with pointers to the input graph representation of the code which specify the input locations to edit.
Although related, it is orthogonal to our work of
learning edits with \pretraining.

\MyPara{Large \Pretrained Models for Code}
Motivated by the success of large \pretrained models for many NLP
tasks, domain-specific models that are \pretrained on source code and
technical text have emerged, including
CodeBERT~\cite{feng2020codebert},
GraphCodeBERT~\cite{guo2020graphcodebert},
CodeGPT-2~\cite{lu2021codexglue}, CodeT5~\cite{wang2021codet5},
PLBART~\cite{ahmad-etal-2021-unified}, PyMT5~\cite{clement2020pymt5},
SynCoBERT~\cite{wang2021syncobert}, SPT-Code~\cite{niu2022spt},
Codex~\cite{ChenETAL21Codex} and UniXcoder~\cite{GuoETAL22Unixcoder}.
Similar to our approach, GraphCodeBERT, \CodeTf, SynCoBERT, SPT-Code and UniXcoder also designed specialized \pretraining objectives driven by their targeted tasks. 
As we showed in this work, the combination of an edit-based language
model and a standard language model can achieve better performance
than using the standard language model alone.

\MyPara{\Pretrained Models for Code Edits}
Prior work already explored applying \pretrained models, despite not
well-suited, on editing tasks.  \citet{ChakrabortyAndRay21Multi} used
PLBART for code bug fixing, which we compared to in
our work. Similarly, \citet{drain2021generating} further \pretrained BART model on 67K
Java repositories mined from GitHub and \finetuned specifically on the bug fixing dataset~\cite{TufanoETAL19Empirical}.
\citet{wang2021codet5, MastropaoloETAL21Studying} both \pretrained T5 model on CodeSerchNet and used it for bug fixing, which we included as a baseline (CodeT5).
Codex~\cite{ChenETAL21Codex} showed promising performance on editing tasks by specifying the existing code as a prompt and providing an edit instruction to the model.
\citet{tufano2022using} and \citet{Li22codereviewer} both proposed
a transformer-based encoder-decoder model \pretrained on large code reviewer
specific data for code review related tasks including
code change quality estimation, review comment generation and
code refinement.
While they demonstrate impressive performance on various tasks, none of them are fundamentally well-suited for edit tasks.
In this work, we develop \MOEDIT with a novel \pretraining objective for generating edit sequences, which can complement the generation model such as \CodeTf for edit tasks.

\MyPara{Combining Complementary Models}  We used
reranking~\cite{NeubigETAL15Reranking,KrizETAL19Reranking} to combine
complementary models in this work.
Ensembling~\cite{LeClairETAL21Ensemble} is another approach for
combining complementary models for generation tasks, but requires
additional training.  Co-training~\cite{BlumAndMitchell98CoTraining}
and tri-training~\cite{ZhouAndLi05TriTraining} approaches, although
shown to be very effective in combining complementary models, are
designed for classification models rather than generation models.


\section{Conclusion}
\label{sec:conclusion}

In this paper, we present a novel edit-driven \pretraining objective and use it to develop \MOEDIT, a \pretrained language model for software-related editing tasks. \MOEDIT is \pretrained on large amounts of source code and natural language comments to perform edits, and we evaluate this model by 
\finetuning it on three distinct downstream tasks:
\cupdate, \bugfix and \codereview. 
By outperforming task-specific baselines and pure generation baselines across tasks, we demonstrate the suitability of \MOEDIT (and our \pretraining objective) for editing tasks and its generalizability. We additionally find that a pure generation-based model and \MOEDIT can complement one another through simple reranking strategies, which outperform each of the models individually and also achieve new state-of-the-art performance for the three downstream editing tasks that we consider.

\begin{acks}
We thank Nader Al Awar, Yu Liu, Raymond J. Mooney, Aditya Thimmaiah,
Zhiqiang Zang, and the anonymous reviewers for their comments and
feedback.  We acknowledge the Texas Advanced Computing Center (TACC)
at The University of Texas at Austin for providing HPC resources that
have contributed to the research results reported within this paper.
This work is partially supported by the US National Science Foundation
under Grant Nos. CCF-1652517, CCF-2107291, IIS-2107524 and
IIS-2145479.
\end{acks}

\balance

\bibliography{bib}

\clearpage

\end{document}